\documentclass[aps,showpacs]{revtex4}
\usepackage{graphicx}
\begin{document}

\title{The statistical hadronization model approach to
$\sqrt{s_{NN}}=200$ GeV Au-Au collisions: $p_{T}$-spectra fits and
global variable predictions}
\author{Dariusz Prorok}
\email{prorok@ift.uni.wroc.pl} \affiliation{Institute of
Theoretical Physics, University of Wroc{\l}aw,\\ Pl.Maksa Borna 9,
50-204 Wroc{\l}aw, Poland}
\date{November 27, 2006}

\begin{abstract}
Three possible scenarios of the statistical hadronization model
are reexamined with the use of the $p_{T}$ spectra of the PHENIX
and very low $p_{T}$ PHOBOS measurements at $\sqrt{s_{NN}}=200$
GeV. These scenarios are: (\textit{a}) full chemical
non-equilibrium, (\textit{b}) strangeness chemical non-equilibrium
and (\textit{c}) chemical equilibrium. Fits to the spectra are
done within the Cracow single-freeze-out model, which takes into
account both the expansion and resonance decays. Predictions for
spectra of $\phi$, $K(892)^{\ast 0}$ and $\pi^{0}$ are also given.
The global variables like the transverse energy at midrapidity,
the charged particle multiplicity at midrapidity and the total
multiplicity of charged particles are evaluated and their
predicted values agree qualitatively well with the experimental
data. The thorough analysis within this model suggests that the
chemical full non-equilibrium case is the least likely and both
other cases are of the similar likelihood. It is also shown that
if the full chemical non-equilibrium freeze-out took place it
could manifest itself in the enhancement of the
$\pi^{0}$-production at very low transverse momenta.
\end{abstract}

 \maketitle

\section {Introduction}
\label{Intrd}

Since the first run of the Relativistic Heavy Ion Collider (RHIC)
the great amount of data on the hadron production from the hot and
dense fireball created in a collision have been available. In this
paper the application of the statistical hadronization model (SHM)
to the description of the fireball bulk properties
\cite{Rafelski:2004dp} is reexamined with the use of the $p_{T}$
spectra measured by the PHENIX Collaboration at the top RHIC
energy $\sqrt{s_{NN}}=200$ GeV \cite{Adler:2003cb}.

In the SHM the formation process of each particle is described on
the basis of the assumption that the accessible phase space is
fully saturated (maximized). Then the particle yields are
determined by their phase space weight which is given by a
statistical distribution (for a comprehensive review of the model
see \cite{Letessier:2002gp}). The main feature of this model is
that it allows to deviate from the usually presumed chemical
equilibrium of the fireball at the freeze-out. This has been
achieved via the introduction of some new parameters, so-called
phase-space occupancy factors: $\gamma_{q}$ for light quarks and
$\gamma_{s}$ for strange quarks in hadrons. In
Ref.~\cite{Rafelski:2004dp} three possible cases were considered:

\begin{enumerate}
 \item Full chemical non-equilibrium, $\gamma_{q} \neq 1$, $\gamma_{s}
\neq 1$.
 \item Strangeness chemical non-equilibrium (semi-equilibrium),
 $\gamma_{q}= 1$, $\gamma_{s}\neq 1$.
 \item Chemical equilibrium, $\gamma_{q}= 1$, $\gamma_{s}= 1$.
\end{enumerate}

The phase-space occupancy factors $\gamma_{q}$ and $\gamma_{s}$
together with the temperature $T$ and baryon number chemical
potential $\mu_{B}$ comprise the full set of independent
statistical parameters of the model. For all three cases of
chemical non-equilibrium/equilibrium, values of these parameters
have been determined in Ref.~\cite{Rafelski:2004dp}. This was done
for each centrality bin of the PHENIX measurement at
$\sqrt{s_{NN}}=200$ GeV from fits to the PHENIX identified hadron
yields \cite{Adler:2003cb} complemented with $K^{\ast}(892)/K^{-}$
and $\phi/K^{-}$ ratios measured by the STAR Collaboration
\cite{Zhang:2004rj,Adams:2004ep,Adams:2004ux}.

However the possible expansion of the fireball is invisible in
particle yield ratios (a collective flow is able to modify only
the momentum spectra of a measured particle, but not its
multiplicity). Therefore, in addition to the studies of particle
ratios the analysis of the $p_{T}$ spectra is necessary to gain
some quantitative information about the flow. Such analysis was
done for the chemical equilibrium case of the SHM (the third point
of the list above) in Ref.~\cite{Prorok:2005uv}. In the present
paper the similar analysis of the $p_{T}$ spectra will be
performed for both chemical non-equilibrium cases of the SHM (the
first and the second point of the list above).

To describe the flow at the stage of the freeze-out, the
single-freeze-out model of
Refs.~\cite{Broniowski:2001we,Broniowski:2001uk,Broniowski:2002nf}
is applied. The model succeeded in the accurate description of
ratios and $p_{T}$ spectra of particles measured at RHIC. The main
postulate of the model is the simultaneous occurrence of chemical
and thermal freeze-outs, which means that the possible elastic
interactions after the chemical freeze-out are neglected. The
conditions for the freeze-out are expressed by values of two
independent thermal parameters: $T$ and $\mu_{B}$. The second
basic feature of the model is the complete treatment of resonance
decays. This means that the final distribution of a given particle
consists not only of the thermal part but also of contributions
from all possible decays and cascades. Feeding from week decays is
included as well. Since in the original formulation
\cite{Broniowski:2001we,Broniowski:2001uk,Broniowski:2002nf} this
model corresponds to the chemical equilibrium case of the SHM, the
generalization of the single-freeze-out model to chemical
non-equilibrium cases of the SHM will be done in the present
paper.

The global variables like the transverse energy at midrapidity
($dE_{T}/d\eta\vert_{mid}$), the charged particle multiplicity at
midrapidity ($dN_{ch}/d\eta\vert_{mid}$) and the total
multiplicity of charged particles ($N_{ch}$) are also evaluated
for both chemical non-equilibrium cases of the SHM for different
centrality bins of the PHENIX measurements at $\sqrt{s_{NN}}=200$
GeV \cite{Adler:2004zn}. These three variables are independent
observables, which means that they are measured independently of
identified hadron spectroscopy. Since model fits were done to
identified hadron data (particle yield ratios and $p_{T}$ spectra)
and the global variables are calculable in the generalized single
freeze-out model, it was natural to check whether their estimated
values agree with the data. This has proven to be true within $10
\%$ accuracy. It should be stressed also here that the centrality
independence of the total multiplicity of charged particles per
participant pair has been reproduced. The evidence for such
scaling of the total multiplicity was reported by the PHOBOS
Collaboration \cite{Back:2006yw}.

In some sense this work could be understood as an additional test
of the correctness of the determination of the statistical
parameters of the SHM since these parameters enter primordial
distributions of hadrons in the fireball at the freeze-out. Thus
fits of geometric parameters of the generalized single-freeze-out
model are done with the use of the values of the statistical
parameters obtained earlier in Ref.~\cite{Rafelski:2004dp} and
treated as input here. The general conclusion is that the best
quality fits to the $p_{T}$ spectra of identified hadrons are
obtained for the strangeness chemical non-equilibrium case of the
SHM. And what is surprising, in the chemical equilibrium case the
spectra seem to be fitted better then in the scenario with the
full chemical non-equilibrium. Whenever in this paper the term
"strangeness chemical non-equilibrium case" or "full chemical
non-equilibrium case" is used it means the case with its values of
statistical parameters taken from Ref.~\cite{Rafelski:2004dp} and
listed in Table~\ref{Table1}, Sec.~\ref{Finlrhotau}. Additionally,
the spectra of $\phi$ and $K(892)^{\ast 0}$ resonances are
predicted. In this way spectra of each particle species whose
yield was used in determination of the statistical parameters of
the model \cite{Rafelski:2004dp} are calculated in here. Also the
measurement of the low momentum $\pi^{0}$ is proposed as a test,
which could help to ascertain whether the full chemical
non-equilibrium could happen in the fireball at the freeze-out or
could not. Namely, values of $\gamma_{q}$ determined in
Ref.~\cite{Rafelski:2004dp} cause that the predictions for
low-$p_{T}$ $\pi^{0}$ are about $40 \%$ greater in this case then
in semi-equilibrium or equilibrium cases.

\section { The single-freeze-out model and its generalization}
\label{Foundat}

The main assumptions of the model are as follows: (\textit{a}) the
chemical and thermal freeze-outs take place simultaneously,
(\textit{b}) all confirmed resonances up to a mass of $2$ GeV from
the Particle Data Tables \cite{Hagiwara:fs} are taken into
account, (\textit{c}) a freeze-out hypersurface is defined by the
equation

\begin{equation}
\tau = \sqrt{t^{2}-r_{x}^{2}-r_{y}^{2}-r_{z}^{2}}= const \;,
\label{Hypsur}
\end{equation}

\noindent (\textit{d}) the four-velocity of an element of the
freeze-out hypersurface is proportional to its coordinate

\begin{equation}
u^{\mu}={ {x^{\mu}} \over \tau}= {t \over \tau}\; \left(1,{
{r_{x}} \over t},{{r_{y}} \over t},{{r_{z}} \over t}\right) \;,
\label{Velochyp}
\end{equation}

\noindent (\textit{e}) the following parameterization of the
hypersurface is chosen:

\begin{equation}
t= \tau \cosh{\alpha_{\parallel}}\cosh{\alpha_{\perp}},\;\;\;
r_{x}=  \tau \sinh{\alpha_{\perp}}\cos{\phi},\;\;\; r_{y}=  \tau
\sinh{\alpha_{\perp}}\sin{\phi},\;\;\;r_{z}=\tau
\sinh{\alpha_{\parallel}}\cosh{\alpha_{\perp}}, \label{Parahyp}
\end{equation}

\noindent where $\alpha_{\parallel}$ is the rapidity of the
element, $\alpha_{\parallel}= \tanh^{-1}(r_{z}/t)$, and
$\alpha_{\perp}$ controls the transverse radius:

\begin{equation}
\rho= \sqrt{r_{x}^{2}+r_{y}^{2}}= \tau \sinh{\alpha_{\perp}} <
\rho_{max} \;, \label{Transsiz}
\end{equation}

\noindent where the restriction on the transverse size has been
introduced, so $\rho_{max}$ gives the maximal transverse extension
of the gas in the central slice during the freeze-out. This means
that two new parameters of the model have been introduced,
\emph{i.e.} $\tau$ and $\rho_{max}$, which are connected with the
geometry of the freeze-out hypersurface.

From Eq.~(\ref{Hypsur}) one can see that the beginning of the
freeze-out process starts at $t_{f.o.}^{(1)}=\tau$ and $\vec{r}=0$
in the c.m.s., which is also the laboratory frame in the RHIC
case. At this moment the volume of the gas can be estimated as

\begin{equation}
V_{f.o.}^{(1)} = 2\pi\tau\rho_{max}^{2} \;, \label{Volgas}
\end{equation}

\noindent which is simply the volume of a tube with a length
$2\tau$ and a radius $\rho_{max}$ ($2\tau$ is the maximal possible
extension of the gas in the longitudinal direction at
$t_{f.o.}^{(1)}$). In the central slice the freeze-out ceases at
$t_{f.o.}^{(2)}= \sqrt{\tau^{2}+\rho_{max}^{2}}$ and it takes
place at $\rho=\rho_{max}$.

The transverse velocity in the central slice can be expressed as a
function of the transverse radius

\begin{equation}
\beta_{\perp}(\rho)= \tanh{\alpha_{\perp}}= { \rho \over
{\sqrt{\tau^{2}+\rho^{2}}}}\;. \label{Betprof1}
\end{equation}

\noindent The maximum value of $\beta_{\perp}$ called the maximum
transverse-flow parameter (or the surface velocity) is given by

\begin{equation}
\beta_{\perp}^{max}= { \rho_{max} \over
{\sqrt{\tau^{2}+\rho_{max}^{2}}}}= { {\rho_{max}/\tau} \over
{\sqrt{1+(\rho_{max}/\tau)^{2}}}}\;. \label{Betmax}
\end{equation}

The invariant distribution of the measured particles of species
$i$ has the form \cite{Broniowski:2001we,Broniowski:2001uk}

\begin{equation}
{ {dN_{i}} \over {d^{2}p_{T}\;dy} }=\int
p^{\mu}d\sigma_{\mu}\;f_{i}(p \cdot u) \;, \label{Cooper}
\end{equation}

\noindent where $d\sigma_{\mu}$ is the normal vector on a
freeze-out hypersurface, $p \cdot u = p^{\mu}u_{\mu}$ , $u_{\mu}$
is the four-velocity of a fluid element and $f_{i}$ is the final
momentum distribution of the particle in question. The final
distribution means here that $f_{i}$ is the sum of primordial and
simple and sequential decay contributions to the particle
distribution (for details see
\cite{Prorok:2004af,Broniowski:2002nf}).

For the most general case of the chemical non-equilibrium the
primordial momentum distribution of particle species $i$ is given
by

\begin{equation}
f_{i}^{primordial}={ {(2s_{i}+1)}  \over {(2\pi\hbar c)^{3}}}\; {
1 \over {\gamma_{i}^{-1} \exp \left\{ {{ E_{i} - \mu_{i} } \over
T} \right\} + g_{i} } }\;, \label{Distrdef}
\end{equation}

\noindent where $E_{i}= ( m_{i}^{2} + p^{2} )^{1/2}$ and $m_{i}$,
$\mu_{i}$, $s_{i}$ and $g_{i}$ are the mass, chemical potential,
spin and a statistical factor of species $i$ respectively. The
chemical potential $\mu_{i} = B_{i}\mu_{B} + S_{i}\mu_{S} +
I^{i}_{3}\mu_{I_{3}}$, where $B_{i}$, $S_{i}$ and $I^{i}_{3}$ are
the baryon number, strangeness and the third component of the
isospin of the particle species in question, whereas $\mu$'s are
the corresponding overall chemical potentials. The strangeness
chemical potential $\mu_{S}$ is determined from the requirement
that the overall strangeness equals zero. The chemical potential
related to the third component of the isospin, $\mu_{I_{3}}$, is
derived from the constraint that the charge to the net baryon
ratio in the final state is the same as in the colliding nuclei.
It has turned out that $\mu_{I_{3}}$ is negligible at RHIC
($\mid\mu_{I_{3}}\mid \leq 1$ MeV
\cite{Rafelski:2004dp,Broniowski:2002nf}), so it will be omitted
in further considerations. The non-equilibrium factor $\gamma_{i}$
reads

\begin{equation}
\gamma_{i}=\gamma_{q}^{(N_{q}^{i}+N_{\bar{q}}^{i})}
\gamma_{s}^{(N_{s}^{i}+N_{\bar{s}}^{i})}\;, \label{Gamnoneq}
\end{equation}

\noindent where $\gamma_{q(s)}$ is the light (strange) quark phase
space occupancy factor, $N_{q}^{i}$ and $N_{s}^{i}$ are the
numbers of light and strange quarks in the \textit{i}th hadron,
and $N_{\bar{q}}^{i}$ and $N_{\bar{s}}^{i}$ are the numbers of the
corresponding antiquarks in the same hadron.

With the use of Eqs.~(\ref{Velochyp}) and (\ref{Parahyp}), the
invariant distribution (\ref{Cooper}) takes the following form:

\begin{equation}
{ {dN_{i}} \over {d^{2}p_{T}\;dy} } = \int d\sigma\;(p \cdot
u)\;f_{i}(p \cdot u) = \tau^{3}\; \int\limits_{-\infty}^{+\infty}
d\alpha_{\parallel}\;\int\limits_{0}^{\rho_{max}/\tau}\;\sinh{\alpha_{\perp}}
d(\sinh{\alpha_{\perp}})\; \int\limits_{0}^{2\pi} d\xi\;(p \cdot
u) \; f_{i}(p \cdot u) \;, \label{Cooper2}
\end{equation}

\noindent where

\begin{equation}
p \cdot u =
m_{T}\cosh{(\alpha_{\parallel}-y)}\cosh{\alpha_{\perp}}-
p_{T}\cos{\xi}\sinh{\alpha_{\perp}}\;. \label{Peu}
\end{equation}

\section {Transverse energy and charged particle
multiplicity} \label{Etncheta}

The experimentally measured transverse energy is defined as

\begin{equation}
E_{T} = \sum_{i = 1}^{L} \hat{E}_{i} \cdot \sin{\theta_{i}} \;,
\label{Etdef}
\end{equation}

\noindent where $\theta_{i}$ is the polar angle, $\hat{E}_{i}$
denotes $E_{i}-m_{N}$ ($m_{N}$ means the nucleon mass) for
baryons, $E_{i}+m_{N}$ for antibaryons and the total energy
$E_{i}$ for all other particles, and the sum is taken over all $L$
emitted particles \cite{Adler:2004zn}.

The pseudorapidity density of particle species $i$ is given by

\begin{equation}
{ {dN_{i}} \over {d\eta} } = \int d^{2}p_{T}\; {{dy} \over {d\eta}
} \; { {dN_{i}} \over {d^{2}p_{T}\;dy} }= \int d^{2}p_{T}\; {p
\over {E_{i}} } \; { {dN_{i}} \over {d^{2}p_{T}\;dy} }\;.
\label{Partdens}
\end{equation}

\noindent Analogously, the transverse energy pseudorapidity
density for the same species can be written as

\begin{equation}
{ {dE_{T,i}} \over {d\eta} } = \int d^{2}p_{T}\; \hat{E}_{i} \cdot
{{p_{T}} \over p} \; {{dy} \over {d\eta} }\; { {dN_{i}} \over
{d^{2}p_{T}\;dy} }= \int d^{2}p_{T}\;{p_{T}} \; { {\hat{E}_{i}}
\over {E_{i}} }\; { {dN_{i}} \over {d^{2}p_{T}\;dy} }\;.
\label{Etraden}
\end{equation}

\noindent For the quantities at midrapidity one has (in the
c.m.s., which is the RHIC case)

\begin{equation}
{ {dN_{i}} \over {d\eta} }\;\Big\vert_{mid}= \int d^{2}p_{T}\; {
{p_{T}} \over {m_{T}} } \;{ {dN_{i}} \over {d^{2}p_{T}\;dy} }\; ,
\label{Partdenmid}
\end{equation}

\begin{equation}
{ {dE_{T,i}} \over {d\eta} }\;\Big\vert_{mid} = \cases{ \int
d^{2}p_{T}\;{p_{T}} \; { {m_{T}-m_{N}} \over {m_{T}} }\; {
{dN_{i}} \over {d^{2}p_{T}\;dy} }\;, i=baryon
 \cr \cr \int
d^{2}p_{T}\;{p_{T}}\; { {m_{T}+m_{N}} \over {m_{T}} } \;{ {dN_{i}}
\over {d^{2}p_{T}\;dy} }\;, i=antibaryon
 \cr \cr \int d^{2}p_{T}\;{p_{T}} \;{ {dN_{i}} \over
{d^{2}p_{T}\;dy} }, i=others  \;.} \label{Etdenmid}
\end{equation}

The overall charged particle and transverse energy densities can
be expressed as

\begin{equation}
{ {dN_{ch}} \over {d\eta} }\;\Big\vert_{mid}= \sum_{i \in B} {
{dN_{i}} \over {d\eta} }\;\Big\vert_{mid}\;, \label{Nchall}
\end{equation}

\begin{equation}
{ {dE_{T}} \over {d\eta} }\;\Big\vert_{mid}= \sum_{i \in A} {
{dE_{T,i}} \over {d\eta} }\;\Big\vert_{mid} \;, \label{Etall}
\end{equation}

\noindent where $A$ and $B$ ($B \subset A$) denote sets of species
of finally detected particles, namely the set of charged particles
$B=\{\pi^{+},\; \pi^{-},\; K^{+},\; K^{-},\; p,\; \bar{p}\}$,
whereas $A$ also includes photons, $K_{L}^{0},\; n$ and
$\bar{n}\;$ \cite{Adcox:2001ry}.

The total multiplicity of particle species $i$ can be also derived
(for the more formal proof see \cite{Broniowski:2002nf})

\begin{eqnarray}
N_{i} &=& \int d^{2}p_{T}\;dy\;{{dN_{i}} \over {d^{2}p_{T}\;dy}}=
\int d^{2}p_{T}\;dy \int p^{\mu}d\sigma_{\mu}\;f_{i}(p \cdot u) =
\int d\sigma \int d^{2}p_{T}\;dy\;(p \cdot u)\;f_{i}(p \cdot u)
\cr \cr && = \int d\sigma \int { d^{3}\vec{p} \over E}\;(p \cdot
u)\;f_{i}(p \cdot u)= \int d\sigma \;
n_{i}(T,\mu_{B},\gamma_{s},\gamma_{q}) =
n_{i}(T,\mu_{B},\gamma_{s},\gamma_{q}) \int d\sigma \;,
\label{Totmult}
\end{eqnarray}

\noindent for any expansion satisfying the condition
$d\sigma_{\mu} \sim u_{\mu}$ on a freeze-out hypersurface and if
the local statistical parameters are constant on this hypersurface
(in the present model both conditions are fulfilled). Note that
the density of particle species $i$, $n_{i}$, includes thermal and
decay contributions. In practise the rapidity of the fluid element
$\alpha_{\parallel}$ should not be unlimited but should have its
maximal value $\alpha_{\parallel}^{max}$. Otherwise, the
hypersurface volume and the total charged particle multiplicity
would be infinite. Then one can express the hypersurface volume as

\begin{equation}
\int d\sigma = \tau^{3}\;
\int\limits_{-\alpha_{\parallel}^{max}}^{+\alpha_{\parallel}^{max}}
d\alpha_{\parallel}\;\int\limits_{0}^{\rho_{max}/\tau}\;\sinh{\alpha_{\perp}}\;
d(\sinh{\alpha_{\perp}})\; \int\limits_{0}^{2\pi} d\xi =
2\pi\;\alpha_{\parallel}^{max} \tau \rho_{max}^{2}
\;.\label{Hypvolum}
\end{equation}

\noindent Finally, the total multiplicity of charged particles can
be obtained:

\begin{equation}
N_{ch} = 2\pi\;\alpha_{\parallel}^{max} \tau \rho_{max}^{2}
\sum_{i \in B} n_{i}(T,\mu_{B},\gamma_{s},\gamma_{q}) =
2\pi\;\alpha_{\parallel}^{max} \tau \rho_{max}^{2}\;
n_{ch}(T,\mu_{B},\gamma_{s},\gamma_{q}) \;.\label{Totcharged}
\end{equation}

For $\alpha_{\parallel}^{max}$ the following reasonable assumption
has been made: it is equal to the rapidity of leading baryons
after the collision. This means that the fluid which has been
created in the central rapidity region (CRR) could not move faster
in the longitudinal direction then fragments of a target or a
projectile after the collision. Therefore
$\alpha_{\parallel}^{max}$ should depend on the centrality of the
collision, since the more central the collision is, the higher
degree of the stopping of the initial baryons ought to happen in
principle. There are two limiting cases, the maximum stopping
happens for the most central collision whereas if the centrality
approaches $100 \%$ the stopping disappears. Assuming additionally
that $\alpha_{\parallel}^{max}$ is a linear function of the
centrality, the following parametrization can be derived (for
details see Ref.~\cite{Prorok:2005uv}):

\begin{equation}
\alpha_{\parallel}^{max}(c) = y_{p} - { \langle \delta y \rangle
\over 0.975 } \cdot (1-c) \;,\label{Alfpmax}
\end{equation}

\noindent where $y_{p}$ is the projectile rapidity, $\langle
\delta y \rangle$ the average rapidity loss and $c$ is a
fractional number representing the middle of a given centrality
bin, \emph{i.e.} $c=0.025$ for the $0-5 \%$ centrality bin,
$c=0.075$ for the $5-10 \%$ centrality bin, etc.. The BRAHMS
Collaboration reports $\langle \delta y \rangle = 2.05$ for the $5
\%$ most central collisions at $\sqrt{s_{NN}}= 200$ GeV ($y_{p} =
5.36$) \cite{Bearden:2003hx}.

\section {Results} \label{Finl}

\subsection {Determination of geometric parameters} \label{Finlrhotau}

\begin{table}
\caption{\label{Table1} Values of the geometric parameters of the
model for various centrality bins fitted with the use of the
PHENIX final data for the $p_{T}$ spectra of identified charged
hadrons \protect\cite{Adler:2003cb}, NDF=124. For bins with
footnotes \textit{a} and \textit{b} the data are from
Ref.~\protect\cite{Adler:2004hv}. Values of the statistical
parameters are taken from \cite{Rafelski:2004dp}. The last three
rows show the results of fits to the set of data which include the
PHENIX data and low-$p_{T}$ $\pi^{+}$ and $\pi^{-}$ data taken as
a half of the PHOBOS data for $(\pi^{+}+\pi^{-})$
\cite{Back:2004zx}, here NDF=132. }
\begin{ruledtabular}
\begin{tabular}{cccccccccccc} \hline Centrality & $N_{part}$ &
 $T$ & $\mu_{B}$
& $\gamma_{s}$ & $\gamma_{q}$ & $\rho_{max}$ & $\tau$ &
$\beta_{\perp}^{max}$ & $V_{f.o.}^{(1)}$ & $t_{f.o.}^{(2)}$ &
$\chi^{2}$/NDF
\\
$[\%]$ & & [MeV] & [MeV] & & & [fm] & [fm] & & [fm$^{3}$] & [fm] &
\\
\hline 0-5 & 351.4 & 141.1 & 25.67 & 2.430 & 1.613 & 7.24$\pm$0.09
& 6.61$\pm$0.06 & 0.74 & 2177.5 & 9.8 & 0.74
\\
5-10 & 299.0 & 141.4 & 24.52 & 2.367 & 1.61169 & 6.82$\pm$0.08 &
6.17$\pm$0.06 & 0.74 & 1804.4 & 9.2 & 0.73
\\
0-10\footnotemark[1] & 325.2 & 141.25 & 25.095 & 2.3985 & 1.6125 &
7.03$\pm$0.08 & 6.39$\pm$0.06 & 0.74 & 1985.8 & 9.5 & 0.80
\\
10-15 & 253.9 & 141.6 & 25.27 & 2.270 & 1.603 & 6.46$\pm$0.08 &
5.81$\pm$0.06 & 0.74 & 1525.6 & 8.7 & 0.72
\\
15-20 & 215.3 & 140.8 & 25.05 & 2.266 & 1.61497 & 6.16$\pm$0.08 &
5.48$\pm$0.05 & 0.75 & 1304.8 & 8.2 & 0.85
\\
20-30 & 166.6 & 141.0 & 26.01 & 2.212 & 1.61387 & 5.58$\pm$0.07 &
4.96$\pm$0.05 & 0.75 & 969.7 & 7.5 & 1.24
\\
30-40 & 114.2 & 142.0 & 25.75 & 2.096 & 1.608 & 4.76$\pm$0.07 &
4.33$\pm$0.05 & 0.74 & 617.0 & 6.4 & 1.64
\\
10-40\footnotemark[2] & 171.8 & 141.35 & 25.52 & 2.211 & 1.61 &
5.80$\pm$0.07 & 5.17$\pm$0.05 & 0.75 & 1093.7 & 7.8 & 1.28
\\
40-50 & 74.4 & 141.7 & 26.14 & 2.003 & 1.605 & 4.06$\pm$0.06 &
3.80$\pm$0.04 & 0.73 & 392.6 & 5.6 & 2.02
\\
50-60 & 45.5 & 141.0 & 24.05 & 1.876 & 1.613 & 3.39$\pm$0.06 &
3.32$\pm$0.04 & 0.71 & 239.1 & 4.7 & 1.96
\\
60-70 & 25.7 & 140.2 & 25.32 & 1.636 & 1.618 & 2.72$\pm$0.05 &
2.86$\pm$0.04 & 0.69 & 133.0 & 3.9 & 2.16
\\
70-80 & 13.4 & 141.7 & 24.24 & 1.026 & 1.299 & 2.40$\pm$0.06 &
2.78$\pm$0.05 & 0.65 & 100.6 & 3.7 & 1.43
\\
\hline 0-5 & 351.4 & 154.6 & 25.04 & 1.231 & 1.0 & 8.35$\pm$0.10 &
8.57$\pm$0.08 & 0.70 & 3752.1 & 12.0 & 0.57
\\
5-10 & 299.0 & 155.2 & 24.73 & 1.186 & 1.0 & 7.84$\pm$0.10 &
7.97$\pm$0.08 & 0.70 & 3077.1 & 11.2 & 0.43
\\
0-10\footnotemark[1] & 325.2 & 154.9 & 24.885 & 1.2085 & 1.0 &
8.10$\pm$0.09 & 8.27$\pm$0.07 & 0.70 & 3407.3 & 11.6 & 0.56
\\
10-15 & 253.9 & 155.5 & 26.29 & 1.169 & 1.0 & 7.37$\pm$0.10 &
7.41$\pm$0.07 & 0.70 & 2527.0 & 10.4 & 0.36
\\
15-20 & 215.3 & 154.6 & 25.68 & 1.147 & 1.0 & 7.07$\pm$0.10 &
7.04$\pm$0.07 & 0.71 & 2212.2 & 10.0 & 0.39
\\
20-30 & 166.6 & 155.2 & 27.18 & 1.121 & 1.0 & 6.37$\pm$0.09 &
6.32$\pm$0.06 & 0.71 & 1609.9 & 9.0 & 0.53
\\
30-40 & 114.2 & 155.7 & 27.21 & 1.080 & 1.0 & 5.47$\pm$0.08 &
5.51$\pm$0.06 & 0.70 & 1036.6 & 7.8 & 0.78
\\
10-40\footnotemark[2] & 171.8 & 155.25 & 26.59 & 1.1293 & 1.0 &
6.65$\pm$0.08 & 6.60$\pm$0.06 & 0.71 & 1833.5 & 9.4 & 0.53
\\
40-50 & 74.4 & 155.5 & 26.74 & 1.018 & 1.0 & 4.65$\pm$0.07 &
4.82$\pm$0.06 & 0.69 & 654.6 & 6.7 & 1.07
\\
50-60 & 45.5 & 152.6 & 21.62 & 0.8906 & 1.0 & 4.06$\pm$0.07 &
4.39$\pm$0.06 & 0.68 & 455.2 & 6.0 & 1.04
\\
60-70 & 25.7 & 152.2 & 26.12 & 0.8076 & 1.0 & 3.22$\pm$0.07 &
3.73$\pm$0.05 & 0.65 & 243.0 & 4.9 & 1.32
\\
70-80 & 13.4 & 148.6 & 23.82 & 0.7163 & 1.0 & 2.59$\pm$0.06 &
3.17$\pm$0.06 & 0.63 & 133.6 & 4.1 & 1.20
\\
80-92 & 6.3 & 150.8 & 28.00 & 0.6788 & 1.0 & 1.92$\pm$0.06 &
2.69$\pm$0.06 & 0.58 & 62.3 & 3.3 & 1.21
\\
\hline 0-5 & 351.4 & 155.2 & 26.4 & 1.0 & 1.0 & 8.46$\pm$0.10 &
8.84$\pm$0.08 & 0.69 & 3973.4 & 12.2 &  0.80
\\
5-10 & 299.0 & 155.2 & 26.4 & 1.0 & 1.0 & 7.99$\pm$0.10 &
8.23$\pm$0.08 & 0.70 & 3302.6 & 11.5 & 0.61
\\
0-10\footnotemark[1] & 325.2 & 155.2 & 26.4 & 1.0 & 1.0 &
8.23$\pm$0.09 & 8.54$\pm$0.07 & 0.69 & 3629.8 & 11.9 & 0.80
\\
10-15 & 253.9 & 155.2 & 26.4 & 1.0 & 1.0 & 7.54$\pm$0.10 &
7.67$\pm$0.08 & 0.70 & 2736.2 & 10.8 & 0.48
\\
15-20 & 215.3 & 155.2 & 26.4 & 1.0 & 1.0 & 7.11$\pm$0.10 &
7.17$\pm$0.07 & 0.70 & 2275.5 & 10.1 & 0.48
\\
20-30 & 166.6 & 155.2 & 26.4 & 1.0 & 1.0 & 6.45$\pm$0.09 &
6.47$\pm$0.07 & 0.71 & 1689.5 & 9.1 & 0.58
\\
30-40 & 114.2 & 155.2 & 26.4 & 1.0 & 1.0 & 5.57$\pm$0.08 &
5.63$\pm$0.06 & 0.70 & 1097.2 & 7.9 & 0.77
\\
10-40\footnotemark[2] & 171.8 & 155.2 & 26.4 & 1.0 & 1.0 &
6.74$\pm$0.08 & 6.76$\pm$0.06 & 0.71 & 1932.3 & 9.6 & 0.64
\\
40-50 & 74.4 & 155.2 & 26.4 & 1.0 & 1.0 & 4.68$\pm$0.07 &
4.85$\pm$0.06 & 0.69 & 669.0 & 6.7 & 1.05
\\
50-60 & 45.5 & 155.2 & 26.4 & 1.0 & 1.0 & 3.83$\pm$0.07 &
4.16$\pm$0.05 & 0.68 & 383.9 & 5.7 & 1.13
\\
60-70 & 25.7 & 155.2 & 26.4 & 1.0 & 1.0 & 2.99$\pm$0.06 &
3.47$\pm$0.05 & 0.65 & 194.3 & 4.6 & 1.41
\\
70-80 & 13.4 & 155.2 & 26.4 & 1.0 & 1.0 & 2.22$\pm$0.06 &
2.78$\pm$0.05 & 0.62 & 86.3 & 3.6 & 1.55
\\
80-92 & 6.3 & 155.2 & 26.4 & 1.0 & 1.0 & 1.71$\pm$0.06 &
2.40$\pm$0.05 & 0.58 & 44.2 & 2.9 & 1.40
\\
\hline \hline 0-15\footnotemark[3] & 303.0 & 141.4 & 25.15 & 2.356
& 1.609 & 6.82$\pm$0.08 & 6.09$\pm$0.05 & 0.75 & 1778.0 & 9.1 &
0.87
\\
\hline 0-15\footnotemark[3] & 303.0 & 155.1 & 25.35 & 1.195 & 1.0
& 7.87$\pm$0.10 & 8.00$\pm$0.07 & 0.70 & 3111.7 & 11.2 & 0.41
\\
\hline 0-15\footnotemark[3] & 303.0 & 155.2 & 26.4 & 1.0 & 1.0 &
8.02$\pm$0.10 & 8.26$\pm$0.08 & 0.70 & 3337.0 & 11.5 & 0.59
\\
\hline
\end{tabular}
\end{ruledtabular}
\footnotetext[1]{Here statistical parameters are the averages of
the parameters listed in two sequential rows above this row.}
\footnotetext[2]{Here statistical parameters are the averages of
the parameters listed in four sequential rows above this row.}
\footnotetext[3]{Here statistical parameters are the averages of
the parameters given for the $0-5 \%$, $5-10 \%$ and $10-15 \%$
centrality classes in the same case of the SHM.}
\end{table}

The determination of parameters of the model proceeds in two
steps. First, statistical parameters $T$, $\mu_{B}$, $\gamma_{q}$
and $\gamma_{s}$ are fitted with the use of the experimental
ratios of hadron multiplicities at midrapidity. This has been
already done in Ref.~\cite{Rafelski:2004dp} for all available
centrality bins of the PHENIX measurements at $\sqrt{s_{NN}}=200$
GeV \cite{Adler:2003cb}. Having put values of these parameters
into the theoretical expression for the invariant distribution,
Eqs.~(\ref{Cooper2}) and (\ref{Peu}), two left parameters
$\rho_{max}$ and $\tau$ can be determined from the simultaneous
fit to the transverse-momentum spectra of $\pi^{\pm}$, $K^{\pm}$,
$p$ and $\bar{p}$. The fits are performed with the help of the
$\chi^{2}$ method.

The final results for the geometric parameters $\rho_{max}$ and
$\tau$ are gathered in Table~\ref{Table1} together with the
corresponding values of $\chi^{2}$/NDF for each centrality class
additionally characterized by the number of participants
$N_{part}$. The results are given for all three cases of the SHM
listed in Sec.~\ref{Intrd} (for comparison the results for the
chemical equilibrium case are repeated from
Ref.~\cite{Prorok:2005uv}). Other physical quantities like the
surface velocity $\beta_{\perp}^{max}$, the volume at the
beginning of the freeze-out $V_{f.o.}^{(1)}$ and the maximal
freeze-out time at the central slice $t_{f.o.}^{(2)}$ are also
given there. Values of $\rho_{max}$ and $\tau$ (therefore also
$V_{f.o.}^{(1)}$ and $t_{f.o.}^{(2)}$) obtained in the case of
full chemical non-equilibrium are substantially lower than
corresponding values in the both other cases. This is because
$\gamma_{s}$ and $\gamma_{q}$ are significantly greater than 1 in
this case, so primordial densities given by Eqs.~(\ref{Distrdef})
and (\ref{Gamnoneq}) are also greater than in both other cases.
And since fits are done to the same spectra, to keep the
normalization unchanged, values of the geometric parameters have
to decrease.

Except the last three rows of Table~\ref{Table1}, all fits have
been done with the use of the $p_{T}$ spectra of identified
charged hadrons measured by the PHENIX Collaboration in
$\sqrt{s_{NN}}=200$ GeV Au-Au collisions
\cite{Adler:2003cb,Adler:2004hv}. Centrality classes with footnote
marks denote two bins for which fitted spectra are taken from
Ref.~\cite{Adler:2004hv}. These are $0-10 \%$ and $10-40 \%$
centrality bins and they are not included in
Ref.~\cite{Adler:2003cb}, so values of the statistical parameters
have not been fitted for them in Ref.~\cite{Rafelski:2004dp}. But
for these bins $\phi$ meson spectra have been reported in
Ref.~\cite{Adler:2004hv}. Thus to make predictions for $\phi$
spectra, values of the statistical parameters have been taken as
the averages of the values fitted for bins which added percent
coverage equals $0-10 \%$ or $10-40 \%$. In the last three rows of
Table~\ref{Table1} there are results of fits to the PHENIX data
complemented with the low-$p_{T}$ data for $\pi^{\pm}$ extracted
from the PHOBOS measurements of $(\pi^{+}+\pi^{-})$
\cite{Back:2004zx}. Since the particle ratio of $\pi^{-}/\pi^{+}
\approx 1$ independently of $p_{T}$ and centrality (see e.g.
Ref.~\cite{Adler:2003cb}), the low-$p_{T}$ values of $\pi^{+}$ and
$\pi^{-}$ spectra  have been taken as one half of
$(\pi^{+}+\pi^{-})$ reported by PHOBOS. However, some modification
of the original PHENIX data \cite{Adler:2003cb} has been done to
match the PHOBOS data conditions. Namely, the PHOBOS measurements
were done for the $15 \%$ most central collisions
($N_{part}=303$), whereas the PHENIX ones for the $0-5 \%$, $5-10
\%$ and $10-15 \%$ centrality bins. Since the treatment of counts
includes the averaging over the number of events in a given
centrality bin and for the same run the number of events in the
$15 \%$ most central bin should be equal to the sum of numbers of
events in the $0-5 \%$, $5-10 \%$ and $10-15 \%$ centrality bins,
the rough approximation of the hypothetic measurement done in the
$0-15 \%$ centrality bin would be the average of the measurements
done in the $0-5 \%$, $5-10 \%$ and $10-15 \%$ centrality bins.
Such averages have been taken as the PHENIX data for the $0-15 \%$
centrality bin. Also values of the statistical parameters taken
for this case are the appropriate averages of the values given for
the $0-5 \%$, $5-10 \%$ and $10-15 \%$ centrality bins.

As it can be seen from the last column of Table~\ref{Table1}, the
best quality fits have been obtained for the strangeness chemical
non-equilibrium case of the SHM. Also fits done in the chemical
equilibrium case are slightly better than those presented for the
full chemical non-equilibrium. This conclusion can be expressed in
an informal quantifiable way by calculating the average of
$\chi^{2}$/NDF for each case of the SHM. So, for the chemical full
non-equilibrium $\langle\chi^{2}$/NDF$\rangle=1.30$, for the
strangeness chemical non-equilibrium
$\langle\chi^{2}$/NDF$\rangle=0.77$ and for the chemical
equilibrium $\langle\chi^{2}$/NDF$\rangle=0.9$. Also the wider
range of centrality fulfils the condition of the statistical
significance, \textit{i.e.} $\chi^{2}$/NDF$ < 1$, in the both
cases of $\gamma_{q}=1$ (up to $40 \%$ of centrality) than in the
case of $\gamma_{q} \neq 1$ (up to $20 \%$ of centrality). Fits
done with the inclusion of the low-$p_{T}$ $\pi^{\pm}$ measured by
PHOBOS have confirmed the above conclusion, as it can be seen in
three last rows of Table~\ref{Table1}.

\begin{figure}
\includegraphics[width=0.42\textwidth]{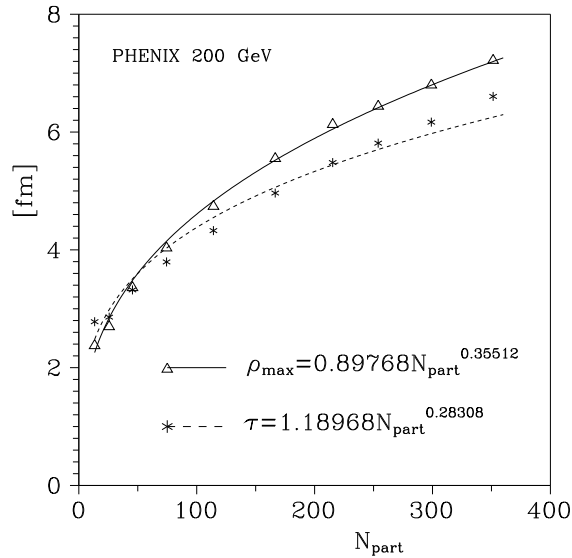}
\caption{\label{Fig.1} Values of the geometric parameters of the
model from the seventh and eighth column of Table~\ref{Table1} for
for the full chemical non-equilibrium case ($\gamma_{s} \neq 1$,
$\gamma_{q} \neq 1$). The lines are the best power approximations.
}
\end{figure}

\begin{figure}
\includegraphics[width=0.42\textwidth]{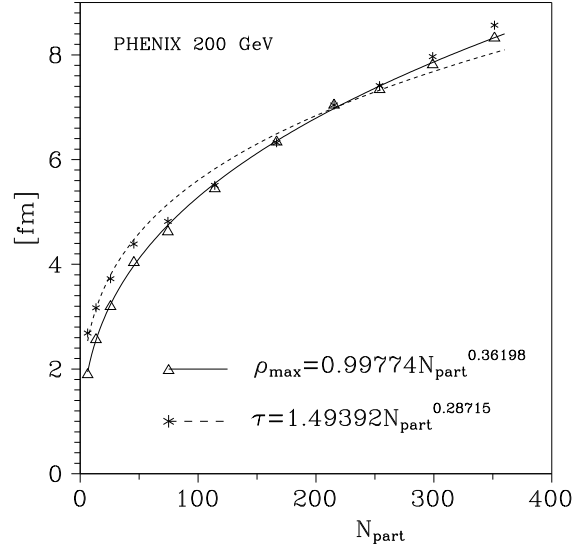}
\caption{\label{Fig.2} Values of the geometric parameters of the
model from the seventh and eighth column of Table~\ref{Table1} for
the strangeness chemical non-equilibrium case ($\gamma_{s} \neq
1$, $\gamma_{q}=1$). The lines are the best power approximations.
}
\end{figure}

Values of the geometric parameters $\rho_{max}$ and $\tau$ from
Table~\ref{Table1} are presented in Figs.\,\ref{Fig.1}-\ref{Fig.2}
as functions of $N_{part}$. Also there the lines of the best power
approximations are depicted,

\begin{equation}
x \sim  N_{part}^{\kappa},\;\;\;\;\;\;\;\; x=\rho_{max},\;\tau,
\label{Scalbeha}
\end{equation}

\noindent with a scaling exponent $\kappa \approx 0.36$ for
$\rho_{max}$ and $\kappa \approx 0.28$ for $\tau$.


\subsection {Identified hadron spectra} \label{Hadrspect}

\begin{figure}
\includegraphics[width=0.42\textwidth]{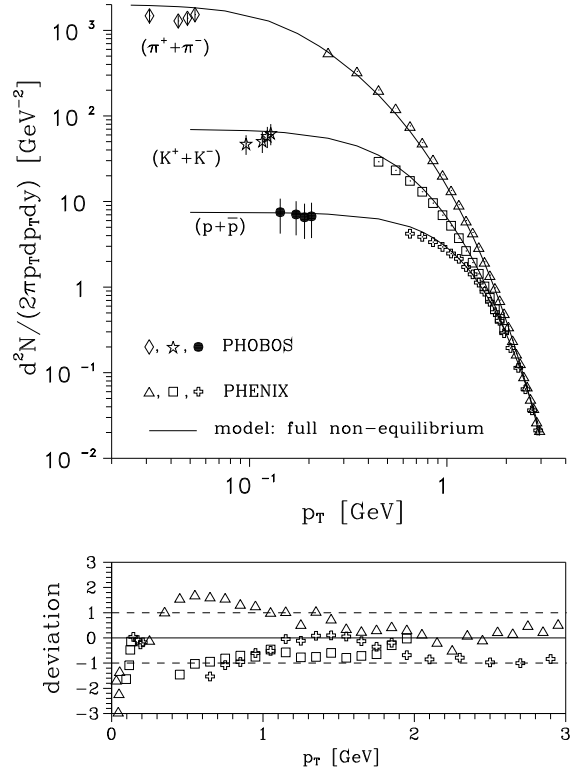}
\caption{\label{Fig.3} The top plot presents invariant yields as a
function of $p_{T}$ for RHIC at $\sqrt{s_{NN}}=200$ GeV. The
PHOBOS data are for the $15 \%$ most central collisions with the
error bars expressed as the sum of the systematic and statistical
uncertainties \protect\cite{Back:2004zx}. The corresponding PHENIX
data \protect\cite{Adler:2003cb} are presented as the averages of
the invariant yields for $0-5 \%$, $5-10 \%$ and $10-15 \%$
centrality bins. For the PHENIX data errors are about $10 \%$ and
are of the size of symbols. Lines are the appropriate predictions
of the single-freeze-out model for the full chemical
non-equilibrium case (fit to the PHENIX data only). The bottom
plot shows a deviation of data to the model,
$(f_{exp}-f_{theo})/\sigma_{exp}$, where $f_{exp(theo)}$ is the
experimental (theoretical) value of the invariant yield at given
$p_{T}$ and $\sigma_{exp}$ is the error of $f_{exp}$. Both PHENIX
and PHOBOS data are denoted by the same symbol for the same
species, \textit{i.e.} triangles are for $(\pi^{+}+\pi^{-})$,
squares for $(K^{+}+K^{-})$ and open crosses for $(p+\bar{p})$.}
\end{figure}

\begin{figure}
\includegraphics[width=0.42\textwidth]{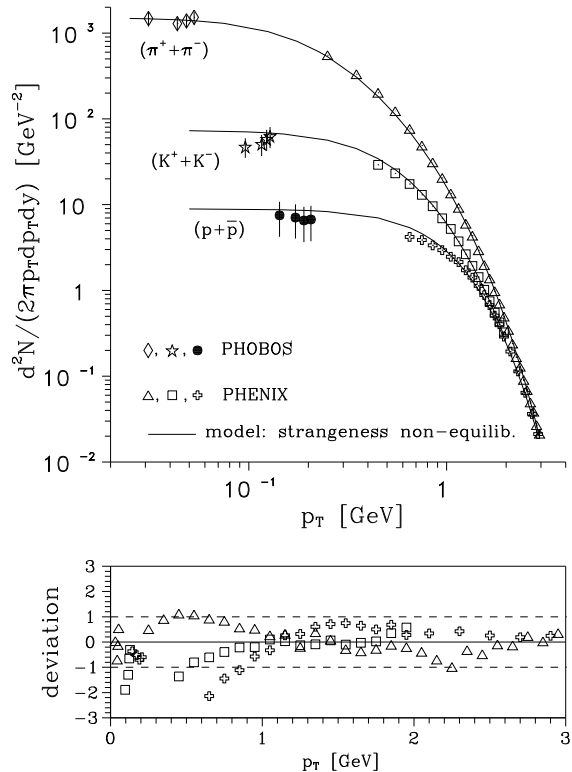}
\caption{\label{Fig.4} Same as Fig.~\ref{Fig.3} but for the
strangeness chemical non-equilibrium case. }
\end{figure}

Having obtained parameters of the model the spectra can be given
with the use of Eqs.~(\ref{Cooper2}) and (\ref{Peu}). In
Figs.~\ref{Fig.3} and \ref{Fig.4} (top plots) the spectra of sums
of negative and positive identified hadrons are depicted. This way
of presentation is chosen to confront the model predictions for
low-$p_{T}$ values of spectra with the PHOBOS experimental data
\cite{Back:2004zx}. Since the PHOBOS data are for the $0-15 \%$
centrality bin, the PHENIX data for this bin have been simulated
in the same way as explained in Sec.~\ref{Finlrhotau}.

\begin{figure}
\includegraphics[width=0.42\textwidth]{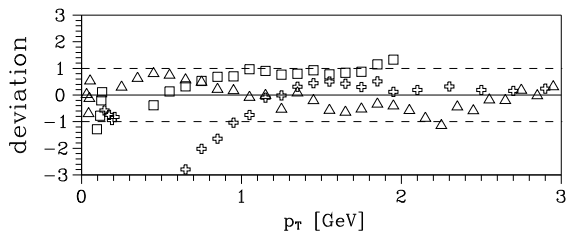}
\caption{\label{Fig.5} Same as the bottom plot of Fig.~\ref{Fig.3}
but for the chemical equilibrium case. }
\end{figure}

In the case of chemical full non-equilibrium, Fig.~\ref{Fig.3},
the low-$p_{T}$ pions are mostly overestimated ($\approx 33 \%$),
after then kaons ($\approx 21 \%$) and the best predictions have
been made for protons and antiprotons ($\approx 5 \%$ above the
data). In the case of chemical strangeness non-equilibrium,
Fig.~\ref{Fig.4}, it is opposite, pions are predicted exactly, but
kaons and protons and antiprotons are overestimated roughly
equally ($\approx 25 \%$). For the chemical equilibrium case (see
Fig. 4 in Ref.~\cite{Prorok:2005uv}, it looks almost the same as
the top plot of Fig.~\ref{Fig.4} here) the situation in the
low-$p_{T}$ range is similar to this in the chemical strangeness
non-equilibrium case, namely pions are in complete agreement with
the data, kaons are $\approx 13 \%$ above and protons and
antiprotons are the most overestimated, $\approx 34 \%$. The above
discussion confirms the conclusion drawn from the comparison of
the values of $\chi^{2}$/NDF - the chemical strangeness
non-equilibrium case seems to work in the best way as far as fits
to the spectra are considered.

To visualize the quality of fits and overall predictions, in the
bottom plots of Figs.~\ref{Fig.3} and \ref{Fig.4} and in
Fig.~\ref{Fig.5} deviations of data to the model are presented.
The deviation is defined as

\begin{equation}
{ {f_{exp}-f_{theo}} \over \sigma_{exp} } \;, \label{Deviatf}
\end{equation}

\noindent where $f_{exp(theo)}$ is the experimental (theoretical)
value of the invariant yield at given $p_{T}$ and $\sigma_{exp}$
is the error of $f_{exp}$. The number of points which are entirely
outside of the $\pm1$ band is the greatest in the case of chemical
full non-equilibrium (21, the most of them corresponds to pions
with all the low-$p_{T}$ sample counted). In the both other cases
this number is the same and equals 8. Also the widest deviation is
in the first case, it reaches -2.9, whereas in the case of
chemical strangeness non-equilibrium the farthest point is -2.1
and in the chemical equilibrium case -2.8.

\begin{figure}
\includegraphics[width=0.42\textwidth]{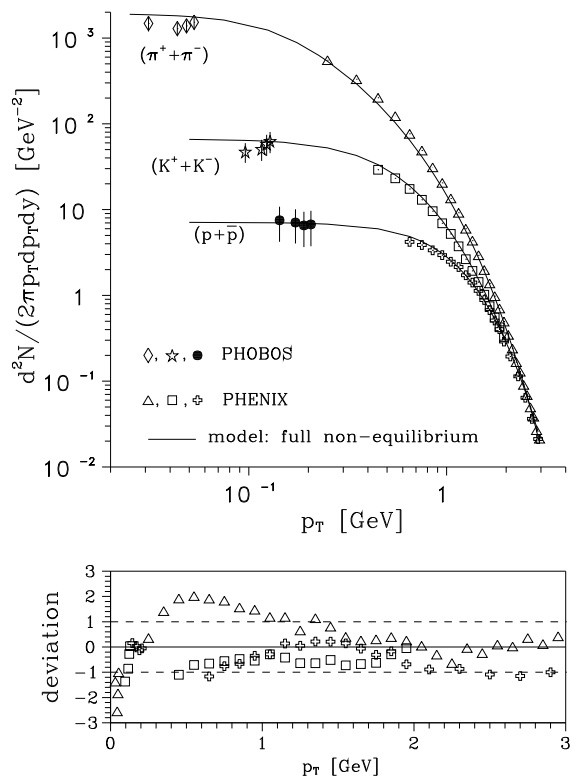}
\caption{\label{Fig.6} Same as Fig.~\ref{Fig.3} but for the
simultaneous fit to the PHENIX and low-$p_{T}$ $\pi^{\pm}$ PHOBOS
data.}
\end{figure}
\begin{figure}
\includegraphics[width=0.42\textwidth]{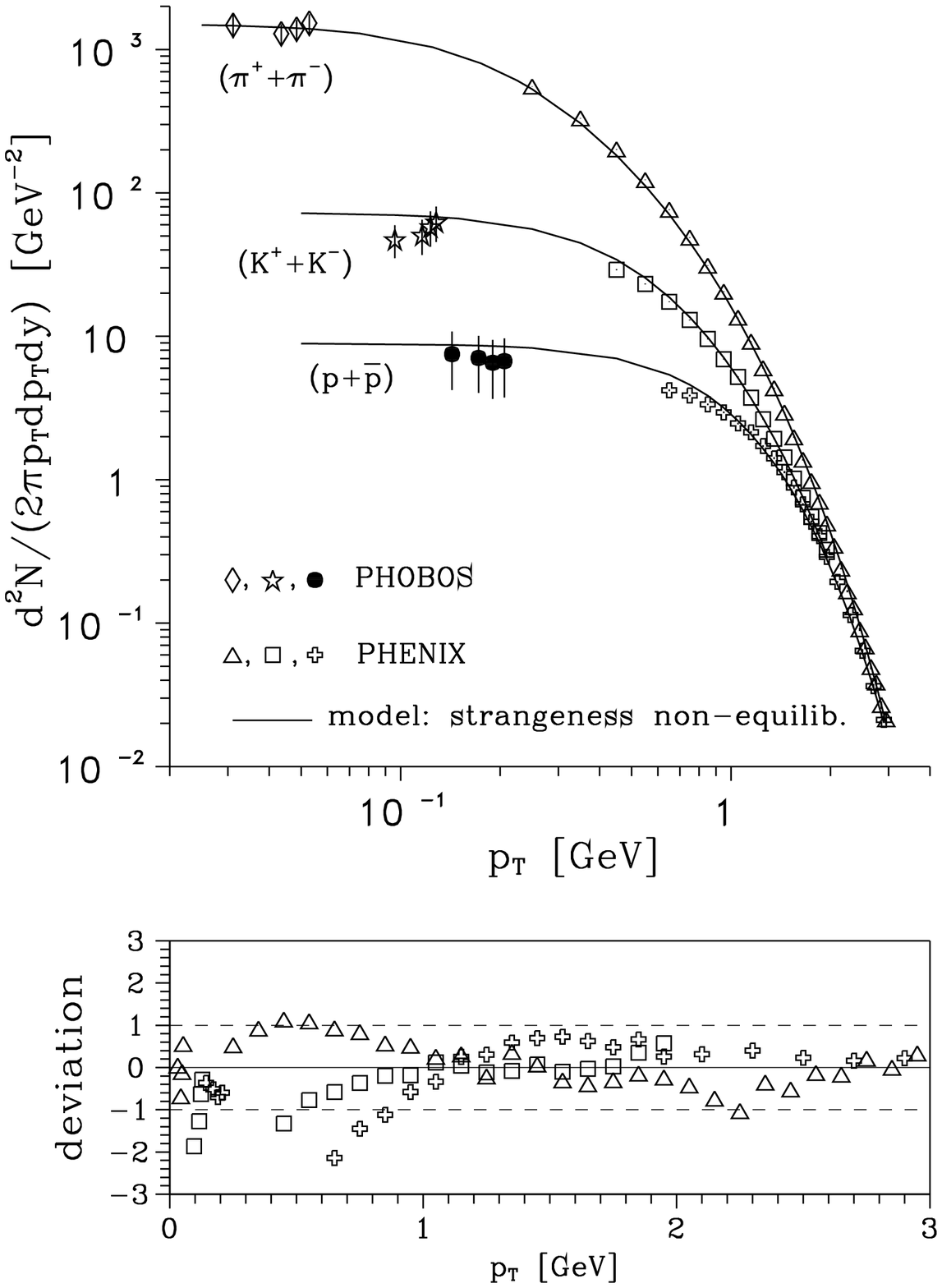}
\caption{\label{Fig.7} Same as Fig.~\ref{Fig.4} but for the
simultaneous fit to the PHENIX and low-$p_{T}$ $\pi^{\pm}$ PHOBOS
data. }
\end{figure}
\begin{figure}
\includegraphics[width=0.42\textwidth]{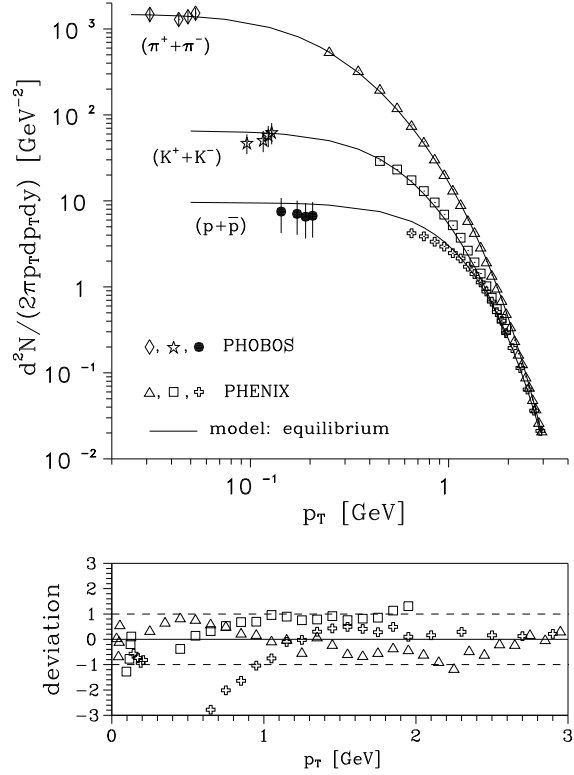}
\caption{\label{Fig.8} Same as Fig.~\ref{Fig.7} but for the
chemical equilibrium case. }
\end{figure}

To investigate into this problem from the other side, fits for the
$15 \%$ most central bin have been done with the inclusion of the
low-$p_{T}$ $\pi^{\pm}$ taken from the PHOBOS data
\cite{Back:2004zx}. The data give values of $(h^{+}+h^{-})$
spectra of identified hadrons ($h=\pi,\;K,\;p$) at very low
$p_{T}$. But only for pions the particle ratio of $h^{-}/h^{+}
\approx 1$ independently of $p_{T}$ and centrality at RHIC (see
e.g. Ref.~\cite{Adler:2003cb}). Thus values of $\pi^{\pm}$ spectra
for very low $p_{T}$ are taken as one half of $(\pi^{+}+\pi^{-})$
reported by PHOBOS \cite{Back:2004zx}. The results of fits have
been gathered in the last three rows of Table~\ref{Table1}. The
corresponding spectra are presented in the top plots of
Figs.~\ref{Fig.6}-\ref{Fig.8}. In the bottom plots of
Figs.~\ref{Fig.6}-\ref{Fig.8} deviations of data to the model are
depicted. Deviation figures shows explicitly what is expressed by
the values of $\chi^{2}$/NDF given in Table~\ref{Table1} - in the
chemical strangeness non-equilibrium case spectra are fitted much
better then in both other cases. The number of points outside the
$\pm1$ band equals 9 in this case and the farthest one is at -2.1
(see the bottom plot of Fig.~\ref{Fig.7}). In the chemical full
non-equilibrium case this number is 19 and the farthest point is
at -2.5 (see the bottom plot of Fig.~\ref{Fig.6}). In the chemical
equilibrium case the number of points outside the $\pm1$ band is
8, but one of them reaches the value -2.8 (see the bottom plot of
Fig.~\ref{Fig.8}).

From what has been explained so far one can see that the chemical
full non-equilibrium freeze-out seems to be less likely in
comparison with semi-equilibrium and equilibrium cases. And if
$\gamma_{q} = 1$ indeed, both last cases will be practically
undistinguishable, however semi-equilibrium will be in favor.

\subsection {$\phi$ and $K(892)^{\ast 0}$ spectra} \label{Phikstar}

\begin{figure}
\includegraphics[width=0.42\textwidth]{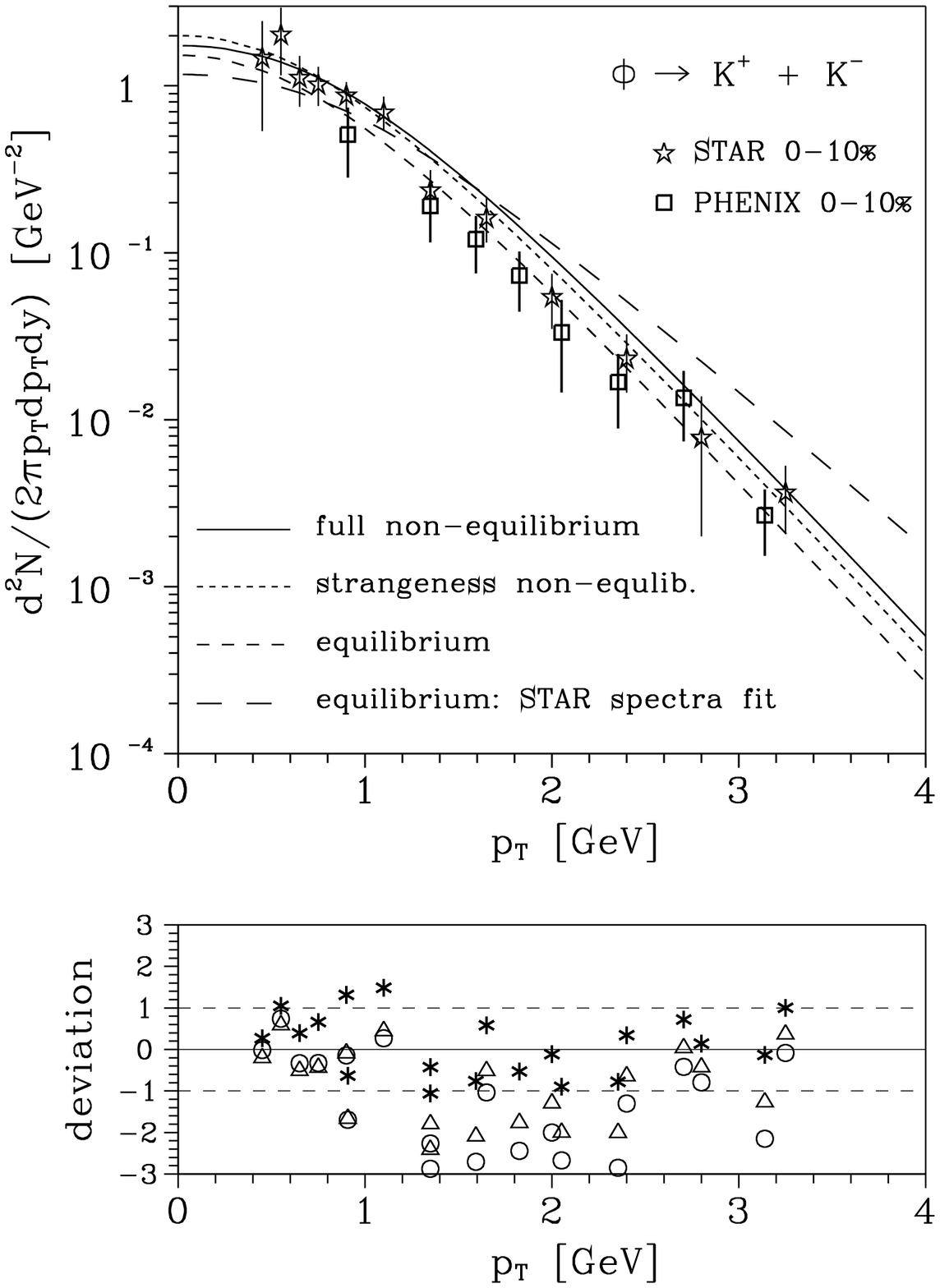}
\caption{\label{Fig.9} The top plot presents invariant yields of
$\phi$ meson measured via the $K^{+}K^{-}$ decay channel as a
function of $p_{T}$ for the $0-10 \%$ centrality bin at
$\sqrt{s_{NN}}=200$ GeV. Data are from
Refs.~\protect\cite{Adams:2004ux} (STAR) and
\protect\cite{Adler:2004hv} (PHENIX). The bottom plot shows a
deviation of the data to the model predictions based on fits to
the PHENIX spectra: chemical full non-equlibrium (circles),
strangeness non-equilibrium (triangles) and chemical equilibrium
(asterisks).}
\end{figure}
\begin{figure}
\includegraphics[width=0.42\textwidth]{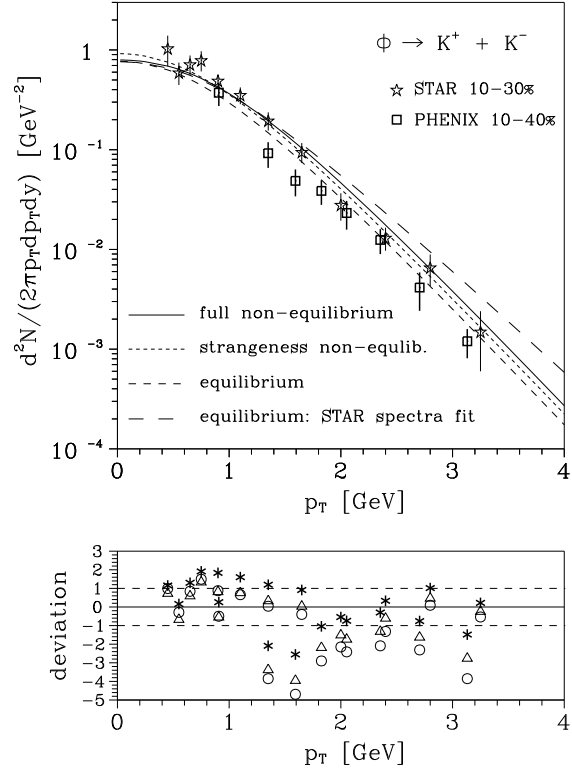}
\caption{\label{Fig.10} Same as Fig.~\ref{Fig.9} but for the
$10-40 \%$ centrality bin. Note that STAR data are for the $10-30
\%$ centrality bin. }
\end{figure}
\begin{figure}
\includegraphics[width=0.42\textwidth]{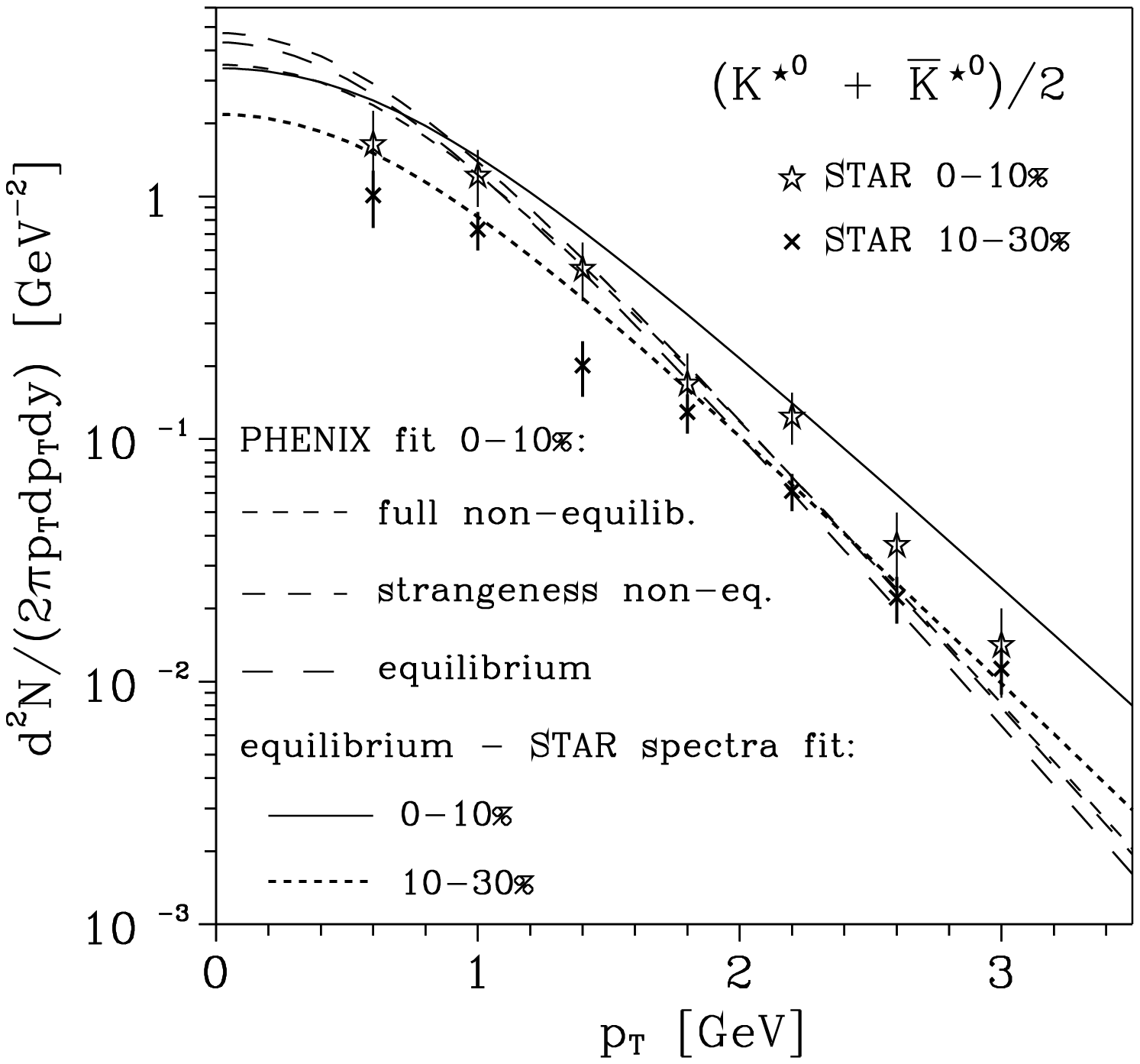}
\caption{\label{Fig.11} The invariant yields of $(K^{\ast
0}+\bar{K}^{\ast 0})/2$ as a function of $p_{T}$ for $0-10 \%$ and
$10-30 \%$ centrality bins at $\sqrt{s_{NN}}=200$ GeV. Data are
from Ref.~\protect\cite{Adams:2004ep}}
\end{figure}

In this subsection the predictions for the spectra of $\phi$ and
$K(892)^{\ast 0}$ resonances will be discussed. This is an
interesting point since the yields of these resonances measured by
the STAR Collaboration
\cite{Zhang:2004rj,Adams:2004ep,Adams:2004ux} were used (with the
basic yields of the identified hadrons measured by the PHENIX
Collaboration \cite{Adler:2003cb}) to fit the statistical
parameters of the model \cite{Rafelski:2004dp}. In the fitting
procedure presented here (to obtain the geometric parameters of
the model $\rho_{max}$ and $\tau$, see Sec.~\ref{Finlrhotau}),
identified hadron spectra measured by PHENIX \cite{Adler:2003cb}
have been explored. So the main source of the data used to test
the SHM here and in Ref.~\cite{Rafelski:2004dp} is the PHENIX
measurement at $\sqrt{s_{NN}}=200$ GeV. But predictions of the
model should be compared with both PHENIX and STAR data, since the
STAR data on $K^{\ast}(892)/K^{-}$ and $\phi/K^{-}$ ratios were
also used in fits of the statistical parameters in
Ref.~\cite{Rafelski:2004dp}. However, one should keep in mind,
when the $\phi$ spectra are discussed, that the $\approx 70 \%$
difference has been found between $\phi$ yields at midrapidity
measured by STAR \cite{Adams:2004ux} and PHENIX
\cite{Adler:2004hv} for one common centrality been, $0-10 \%$. The
reason for such behavior is still unknown and this is probably not
a statistical fluctuation of the lowest $m_{T}$ PHENIX point, as
suggested in Ref.~\cite{Rafelski:2004dp}, because when the same
$m_{T}$ range for both the STAR and the PHENIX $\phi$ data is
considered the difference still persists (see Ref. [59] in
Ref.~\cite{Adler:2004hv}).

In Figs.~\ref{Fig.9} and \ref{Fig.10} predictions for the $\phi$
production are presented for $0-10 \%$ and $10-40 \%$ centrality
classes of the PHENIX measurement for all three cases of the SHM
analyzed here. Additionally, the results for the equilibrium case
but such that the statistical and geometric parameters of the
model are fitted to the STAR data only are also depicted. This is
the case considered in Ref.~\cite{Prorok:2005uv}: the statistical
parameters ($T = 160.0$ MeV, $\mu_{B} = 24.0$ MeV) are fitted to
the STAR particle yield ratios \cite{Barannikova:2005rw} and the
geometric parameters to the $p_{T}$ spectra of identified hadrons
delivered by the STAR Collaboration in Ref.~\cite{Adams:2003xp}.
Again, since the STAR identified hadron spectra
\cite{Adams:2003xp} are for different centrality classes than the
STAR $\phi$-spectra \cite{Adams:2004ux}, the values of geometric
parameters for $0-10 \%$ and $10-30 \%$ centrality bins explored
by STAR in $\phi$ meson measurements are the averages of the
values fitted in Ref.~\cite{Prorok:2005uv} for bins which added
percent coverage equals $0-10 \%$ and $10-30 \%$ respectively.
This gives $\rho_{max}=8.81$ fm, $\tau=6.98$ fm for the $0-10 \%$
centrality bin and $\rho_{max}=7.035$ fm, $\tau=6.095$ fm for the
$10-30 \%$ centrality bin. Results corresponding to these two
equilibrium (STAR) cases are presented as long-dashed lines in
Figs.~\ref{Fig.9} and \ref{Fig.10}. Also both the PHENIX and the
STAR data are depicted in these figures. Note that the STAR second
bin is $10-30 \%$, whereas the second PHENIX bin is $10-40 \%$
(see Fig.~\ref{Fig.10}). Generally, as one can see in
Figs.~\ref{Fig.9} and \ref{Fig.10}, all three cases of the SHM
agree qualitatively with both the PHENIX and the STAR data when
the predictions are based on the fits to the PHENIX spectra
(solid, short-dashed and dashed lines). When the predictions are
based on the fit to the STAR spectra (long-dashed lines) they
agree with the STAR data only up to the intermediate transverse
momentum range and overestimate the high $p_{T}$ data. It is not
clear why this happens, but within the PHENIX experiment the
picture is consistent.

Since Figs.~\ref{Fig.9} and \ref{Fig.10} are in the logarithmic
scale, to see quantitative agreement or disagreement of each SHM
case, one should go somehow to the linear scale. This has been
done in the bottom plots of Figs.~\ref{Fig.9} and \ref{Fig.10},
where deviations of both the PHENIX and the STAR data to the model
are depicted for all three cases of the SHM considered here. It is
clearly seen that predictions in the equilibrium case are
substantially better than in the both non-equilibrium cases. In
fact, the both non-equilibrium cases are practically ruled out,
however the strangeness non-equilibrium case seems to behave
slightly better than the full non-equilibrium case.

In Fig.~\ref{Fig.11} results for $(K^{\ast 0}+\bar{K}^{\ast 0})/2$
spectra are presented together with the STAR data for $0-10 \%$
and $10-30 \%$ centrality classes \cite{Adams:2004ep}. This figure
is very instructive since it explicitly shows that the data from
different collaborations should not be mixed in any fitting
procedure. As one can see predictions based on fitting to the
PHENIX data for the $0-10 \%$ centrality class (\textit{cf}.
Table~\ref{Table1}) are very similar to each other and miss the
STAR data on the $(K^{\ast 0}+\bar{K}^{\ast 0})/2$ production
mainly because of the different slope. In fact, these predictions
are above the level of the STAR data for the $0-10 \%$ centrality
class in the low transverse momenta but they are below even the
STAR data for the $10-30 \%$ centrality class in the high
transverse momenta. Additionally, the results for the equilibrium
case but such that the statistical and geometric parameters of the
model are fitted to the STAR data only are also depicted. This is
the case considered in Ref.~\cite{Prorok:2005uv}: the statistical
parameters ($T = 160.0$ MeV, $\mu_{B} = 24.0$ MeV) are fitted to
the STAR particle yield ratios \cite{Barannikova:2005rw} and the
geometric parameters to the $p_{T}$ spectra of identified hadrons
delivered by the STAR Collaboration in Ref.~\cite{Adams:2003xp}.
Again, since the STAR identified hadron spectra
\cite{Adams:2003xp} are for different centrality classes than the
STAR $(K^{\ast 0}+\bar{K}^{\ast 0})/2$ spectra
\cite{Adams:2004ep}, the values of geometric parameters for $0-10
\%$ and $10-30 \%$ centrality bins explored by STAR in $(K^{\ast
0}+\bar{K}^{\ast 0})/2$ measurements are the averages of the
values fitted in Ref.~\cite{Prorok:2005uv} for bins which added
percent coverage equals $0-10 \%$ and $10-30 \%$ respectively.
This gives $\rho_{max}=8.81$ fm, $\tau=6.98$ fm for the $0-10 \%$
centrality bin and $\rho_{max}=7.035$ fm, $\tau=6.095$ fm for the
$10-30 \%$ centrality bin. Results corresponding to these two
equilibrium (STAR) cases are presented as solid and
shortest-dashed lines in Figs.~\ref{Fig.11}. In fact some
overestimation in normalization can be seen, mostly in the case of
the $0-10 \%$ centrality bin, but slopes are correct.

\subsection {$\pi^{0}$ spectra} \label{Pizero}

\begin{figure}
\includegraphics[width=0.42\textwidth]{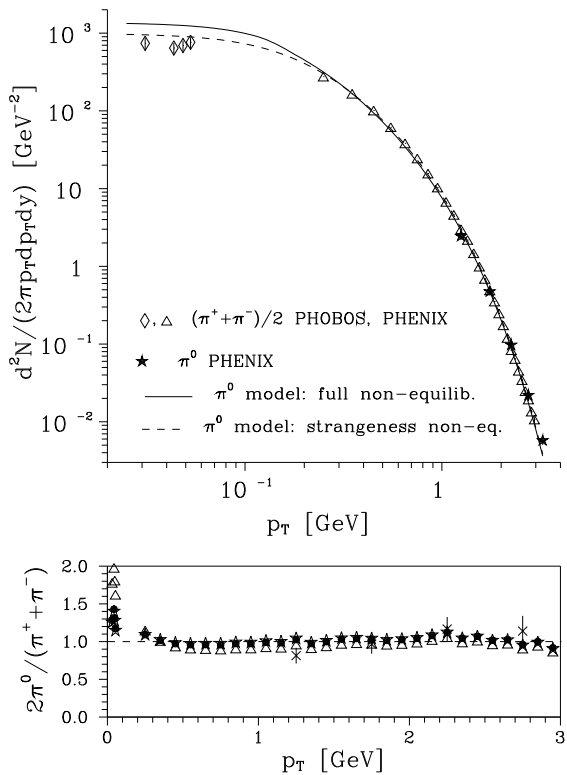}
\caption{\label{Fig.12} The top plot presents invariant yields of
$\pi^{0}$ in comparison with the half of $(\pi^{+}+\pi^{-})$
yields for RHIC at $\sqrt{s_{NN}}=200$ GeV. The PHOBOS data
(diamonds) are for the $15 \%$ most central collisions
\protect\cite{Back:2004zx}. The corresponding PHENIX data
\protect\cite{Adler:2003cb} (triangles) are presented as the
averages of the invariant yields for the $0-5 \%$, $5-10 \%$ and
$10-15 \%$ centrality bins. The PHENIX data for $\pi^{0}$ (stars)
are for the $10 \%$ most central bin \protect\cite{Adler:2003qi}.
For both PHENIX data errors are about $10 \%$ and are of the size
of symbols. Lines are the appropriate predictions of the
single-freeze-out model. The bottom plot shows a ratio of
predicted $\pi^{0}$ to the half of measured $(\pi^{+}+\pi^{-})$
for the full chemical non-equilibrium case (triangles), the
strangeness chemical non-equilibrium case (dots) and the chemical
equilibrium case (open stars). Also values of the ratio calculated
with the use of the experimental data are depicted (crosses). }
\end{figure}

The occupancy factor $\gamma_{q}$, when differs from one, could
influence the $\pi^{0}$ spectra strongly. This is because for
$\pi^{0}$ $N_{q}=N_{\bar{q}}=1$ and $N_{s}=N_{\bar{s}}=0$. Then in
the primordial distribution of $\pi^{0}$ one has (see
Eq.~(\ref{Distrdef}))

\begin{equation}
\gamma_{\pi^{0}}^{-1} \exp \left\{ { E_{\pi^{0}} \over T} \right\}
= \exp \left\{ {{ E_{\pi^{0}} - \mu_{\pi}} \over T} \right\}\;,
\label{Pizerfact}
\end{equation}

\noindent where the chemical potential of pions is defined as

\begin{equation}
\mu_{\pi} = 2T\ln{\gamma_{q}}\;. \label{Pichempot}
\end{equation}

\noindent It has turned out that in the case of chemical full
non-equilibrium

\begin{equation}
\mu_{\pi} \approx m_{\pi^{0}}\;\;\;(\textrm{but}\;\; \mu_{\pi}\leq
m_{\pi^{0}}\;\; \textrm{always}) \label{Pipoteqmas}
\end{equation}

\noindent for the fitted values of $\gamma_{q}$ and $T$ taken from
Ref.~\cite{Rafelski:2004dp} and listed in Table~\ref{Table1}. This
means that in this case the values of the statistical parameters
happen to hit the critical values for the Bose-Einstein
condensation of neutral pions. This has been already stated by the
authors of Ref.~\cite{Rafelski:2004dp} in
Refs.~\cite{Rafelski:2003ju,Letessier:2005qe}, namely that if
$\gamma_{q}$ is freed from 1 but is kept in the range
$[1,\gamma_{q}^{cr}=e^{m_{\pi^{0}}/2T}]$, it goes to its critical
value $\gamma_{q}^{cr}$ during fitting procedure. If this happened
really, this could enhance the production of $\pi^{0}$'s with very
low $p_{T}$.

Predictions for $\pi^{0}$ spectra are presented in
Fig.~\ref{Fig.12} for two non-equilibrium cases of the SHM. The
$\pi^{0}$ spectrum in the chemical equilibrium case is roughly the
same as that for the chemical strangeness non-equilibrium case, so
it is not depicted. However it is impossible to compare the
predictions for the low $p_{T}$ with the data since the
appropriate data have not been available yet. Thus in
Fig.~\ref{Fig.12} the comparison is done with the half of
$(\pi^{+}+\pi^{-})$ spectrum delivered by PHOBOS for the $15 \%$
most central bin \cite{Back:2004zx} and with the corresponding
spectrum compiled from the PHENIX data \cite{Adler:2003cb}. As it
is shown in the bottom plot of Fig.~\ref{Fig.12}, the experimental
ratio of $2\pi^{0}/(\pi^{+}+\pi^{-}) \approx 1$ in the range of
$p_{T}$ common for $\pi^{\pm}$ \cite{Adler:2003cb} and $\pi^{0}$
\cite{Adler:2003qi} PHENIX measurements at $\sqrt{s_{NN}}=200$
GeV, \textit{i.e.} for $1 < p_{T} < 3$ GeV.

One can see from the top plot of Fig.~\ref{Fig.12} that down to
the $p_{T} \approx 0.2$ GeV all three cases of the SHM predict
roughly the same spectrum of $\pi^{0}$ (the curve for the chemical
equilibrium case is not depicted because, in the logarithmic
scale, it would exactly cover the curve for the chemical
strangeness non-equilibrium case). The difference between
predictions in the chemical full non-equilibrium case and
predictions in both other cases arises at very low transverse
momenta and is about $40 \%$. This can be seen very clearly in the
bottom plot of Fig.~\ref{Fig.12}, where the ratio of predicted
$\pi^{0}$ to the half of measured $(\pi^{+}+\pi^{-})$ is depicted
as a function of $p_{T}$ for all three cases of the SHM. The
enhancement of neutral pions over one half of charged pions is
$\approx 80 \%$ in the case of chemical full non-equilibrium,
whereas for both other cases is $\approx 30 \%$. This suggests
that the measurement of very low $p_{T}$ $\pi^{0}$'s could be
helpful to judge whether $\gamma_{q} \approx \gamma_{q}^{cr}$ (as
is claimed in
Refs.~\cite{Rafelski:2004dp,Rafelski:2003ju,Letessier:2005qe} on
the basis of fits to particle yields/ratios) or $\gamma_{q}=1$.

\subsection {Transverse energy and charged particle multiplicity
estimations} \label{Finletnch}

\begin{figure}
\includegraphics[width=0.42\textwidth]{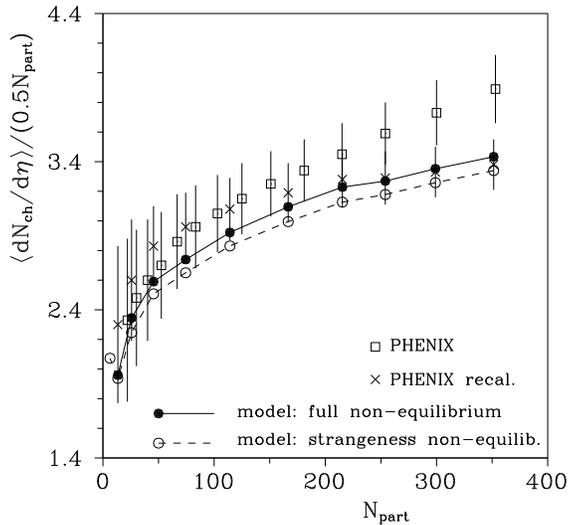}
\caption{\label{Fig.13} $dN_{ch}/d\eta$ per pair of participants
versus $N_{part}$ for RHIC at $\sqrt{s_{NN}}=200$ GeV. The
original PHENIX data are from Ref.~\protect\cite{Adler:2004zn},
whereas the recalculated PHENIX data are from summing up the
integrated charged hadron yields delivered in
Ref.~\protect\cite{Adler:2003cb}. The lines connect the results
and are a guide. }
\end{figure}

The results of numerical estimations of $dN_{ch}/d\eta\vert_{mid}$
divided by the number of participant pairs for various centrality
classes are presented in Fig.~\ref{Fig.13} for RHIC at
$\sqrt{s_{NN}}=200$ GeV. The results are given for two
non-equilibrium cases of the SHM (the estimates in the case of
chemical equilibrium are almost the same as in the chemical
strangeness non-equilibrium case and have been already presented
in Ref.~\cite{Prorok:2005uv}, see Fig.6 therein). Additionally to
the straightforward PHENIX measurements of the charged particle
multiplicity density, the data from the summing up of the
integrated charged hadron yields \cite{Adler:2003cb} are depicted
in these figures, too (these data are called "recalculated", for
more explanations see Ref.~\cite{Prorok:2005uv}). Note that the
recalculated data differ from the direct ones, especially for more
central bins. This has been already noticed by the PHENIX
Collaboration (see backup slides of \cite{Chujo:2002bi}). In
\cite{Adler:2003cb} the feeding of $p(\bar{p})$ from
$\Lambda(\bar{\Lambda})$ decays is excluded. To diminish this
effect, integrated $p$ and $\bar{p}$ yields delivered in
\cite{Adler:2003cb} were corrected to include back the feeding.
The correction was done by the division by a factor 0.65, which is
the rough average of a $p_{T}$-dependent multiplier used by PHENIX
Collaboration (see Fig.4 in \cite{Adler:2003cb} and Eq.(5)
therein).

Generally, in both presented cases the model predictions
underestimate the directly measured (more) as well as recalculated
(less) $dN_{ch}/d\eta\vert_{mid}$. However, the estimates in the
chemical full non-equilibrium case are slightly closer to the data
and for the four most central bins they agree entirely with the
recalculated data points. In the full range of centrality the
predictions agree with the recalculated data within errors in this
case and almost agree within errors in the chemical strangeness
non-equilibrium case. In principle, since the fits of the
geometric parameters of the model have been done to the same
$p_{T}$ spectra here, which were integrated to deliver charged
hadron yields in Ref.~\cite{Adler:2003cb}, the predictions for
$dN_{ch}/d\eta\vert_{mid}$ should agree exactly with the
recalculated data. However, the transverse momentum spectra are
measured in \emph{limited ranges}, so very important low-$p_{T}$
regions are blank in Ref.~\cite{Adler:2003cb}. To obtain
integrated yields some extrapolations below and above the measured
ranges are used. In fact these extrapolations are only analytical
fits, but contributions from regions covered by them account for
about $25-40\%$ of the integrated yields \cite{Adcox:2001mf}.
These extrapolations could differ from the distributions obtained
in the framework of this model and this could be the main source
of the discrepancy between the predictions and the recalculated
data. So the question why the significant underestimation of the
predicted $dN_{ch}/d\eta\vert_{mid}$ with respect to the directly
measured charged particle multiplicity density occurs should be
addressed to the experimentalists rather: why does the directly
measured $dN_{ch}/d\eta\vert_{mid}$ differ substantially from the
sum of the integrated hadron yields for central collisions?

\begin{figure}
\includegraphics[width=0.42\textwidth]{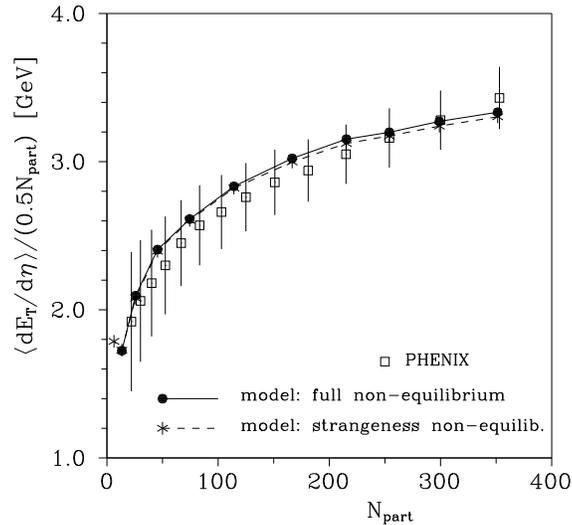}
\caption{\label{Fig.14} $dE_{T}/d\eta$ per pair of participants
versus $N_{part}$ for RHIC at $\sqrt{s_{NN}}=200$ GeV. The PHENIX
data are from Ref.~\protect\cite{Adler:2004zn}. The lines connect
the results and are a guide. }
\end{figure}

The values of $dE_{T}/d\eta\vert_{mid}$ per pair of participants
as a function of participant pairs are shown in Fig.~\ref{Fig.14}
for $\sqrt{s_{NN}}=200$ GeV. The quality of the model predictions
is much better in this case then for $dN_{ch}/d\eta\vert_{mid}$,
they agree with the data almost completely. Note that predictions
in both presented cases are practically the same and do not differ
from the corresponding results in the chemical equilibrium case
(see Fig.7 in Ref.~\cite{Prorok:2005uv}).

\begin{figure}
\includegraphics[width=0.42\textwidth]{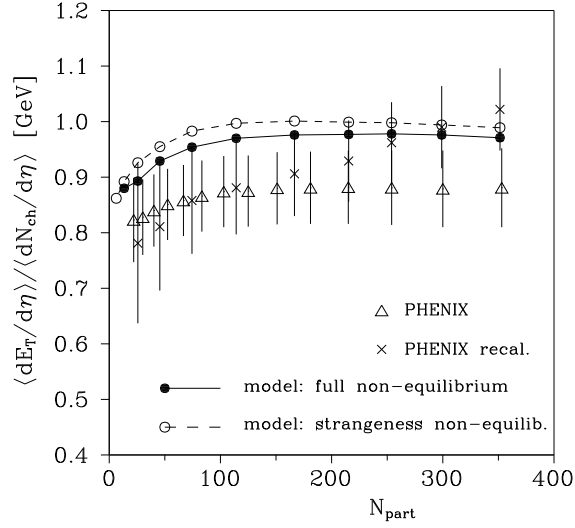}
\caption{\label{Fig.15} $\langle dE_{T}/d\eta\rangle /\langle
dN_{ch}/d\eta\rangle$ versus $N_{part}$ for RHIC at
$\sqrt{s_{NN}}=200$ GeV. The original PHENIX data are from
Ref.~\protect\cite{Adler:2004zn}. The recalculated PHENIX data are
also depicted, here "recalculated" means that the sum of
integrated charged hadron yields \cite{Adler:2003cb} have been
substituted for the denominator in the ratio. The lines connects
the results and are a guide. }
\end{figure}

Values of the ratio $\langle dE_{T}/d\eta\rangle /\langle
dN_{ch}/d\eta\rangle$ as a function of $N_{part}$ are presented in
Fig.~\ref{Fig.15}. Again, as for $dN_{ch}/d\eta\vert_{mid}$, the
values predicted in the chemical full non-equilibrium case are
slightly closer to the data, they agree with the recalculated data
within errors. For the most central bins both sets of predictions
agree with the recalculated data within errors. $\langle
dE_{T}/d\eta\rangle /\langle dN_{ch}/d\eta\rangle$ estimates done
within the chemical equilibrium case are practically the same as
in the strangeness non-equilibrium case (see Fig.10 in
Ref.~\cite{Prorok:2005uv}). As far as the comparison with the
direct data \cite{Adler:2004zn} is concerned, the position of
model predictions is very regular and exactly resembles the
configuration of the data in each case, the estimates are only
shifted up about $10 \%$ as a whole.

\begin{figure}
\includegraphics[width=0.42\textwidth]{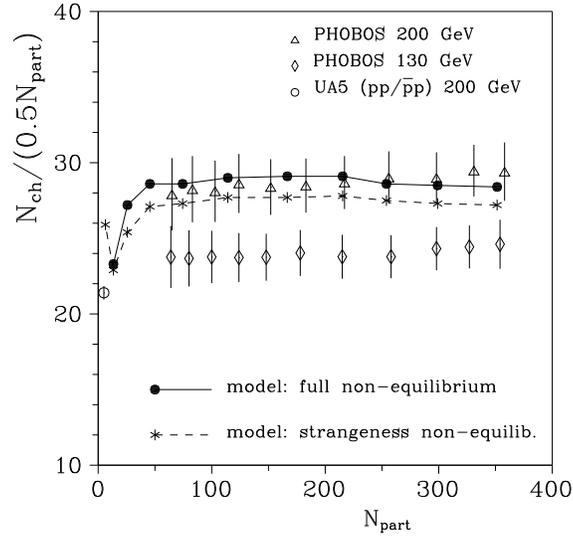}
\caption{\label{Fig.16} $N_{ch}$ per pair of participants versus
$N_{part}$ for RHIC at $\sqrt{s_{NN}}=200$ GeV. The PHOBOS data
are from Ref.~\protect\cite{Back:2006yw} and the $pp/\bar{p}p$
data point of the UA5 measurement is from Fig.39.5 in
Ref.~\protect\cite{Hagiwara:fs}. The lines connect the results and
are a guide. }
\end{figure}

The last discussed global variable is the total multiplicity of
charged particles $N_{ch}$, which can be calculated with the use
of Eqs.~(\ref{Totcharged}) and (\ref{Alfpmax}). The results
presented as the total charged-particle multiplicity per
participating pair versus $N_{part}$ are gathered in
Fig.\,\ref{Fig.16}. The both sets of predictions exhibit almost
ideal centrality independence within the range of the PHOBOS
measurement, \textit{i.e.} $N_{part} \approx 60-360$. Note that in
the chemical full non-equilibrium case also normalization agrees
almost exactly with the data. In the case of chemical strangeness
non-equilibrium the $6 \%$ underestimation has resulted but the
predictions still agree with the data within errors. For the
chemical equilibrium the similar underestimation was obtained
($6.4 \%$, see Fig.12 in Ref.~\cite{Prorok:2005uv}).

The general conclusion which could be drawn from the above
discussion is that predictions for global variables agree pretty
well with the data in each case of the SHM. However, the chemical
full non-equilibrium scenario works slightly better in this
respect.

\section {Conclusions}
\label{Conclud}

The extensive analysis of the RHIC data on the particle production
in Au-Au collisions at $\sqrt{s_{NN}}=200$ GeV has been performed
within three possible scenarios of the statistical hadronization
model. The SHM explored here is the generalized version of the
model of Ref.~\cite{Rafelski:2004dp}. The generalization means the
explicite inclusion of the fireball expansion in a way as proposed
in the single-freeze-out model of
Refs.~\cite{Broniowski:2001we,Broniowski:2001uk,Broniowski:2002nf}.

Generally no definite rejection of any of these scenarios could be
done on the basis of this analysis, since different observables
prefer different scenarios. However, the chemical full
non-equilibrium case seems to be the least likely. This is evident
from the studies of particle spectra of stable charged hadrons,
both from the statistical significance point of view (values of
$\chi^{2}$/NDF in Table~\ref{Table1}) and from the behavior of the
deviation factor (bottom plots of Figs.~\ref{Fig.3}-\ref{Fig.8}).
The $\phi$-spectrum test confirms the above conclusion (see
Sec.~\ref{Phikstar}). On the opposite, the global variable test
prefers the full chemical non-equilibrium scenario and does not
distinguish between strangeness chemical non-equilibrium and
chemical equilibrium cases (see Sec.~\ref{Finletnch}) but the
differences are not significant. The semi-equilibrium and
equilibrium scenarios seem to be of the similar likelihood, even
though the $\phi$-spectrum analysis discredits the strangeness
chemical non-equilibrium case. This is because the fits to spectra
of identified charged hadrons seem to weigh most when the
conclusion is to be drawn from the present studies. The $p_{T}$
spectra of stable charged hadrons comprise the most numerous and
of the highest quality samples of the experimental data. For each
centrality class of the PHENIX measurement at $\sqrt{s_{NN}}=200$
GeV, the $p_{T}$ spectra counts more than 120 points, whereas each
sample of the discussed resonance spectra or of the global
variable has about 10 points. Also the measurement of a stable
charged hadron seems to be more accurate since such a hadron is
measured directly while resonances can be measured only via their
decay products. The last point which supports the relevance of the
fits to identified hadron spectra is that all data used in the
fitting procedure (here particle yields and $p_{T}$ spectra) and
the data which predictions are compared with, should originate
from the same experiment, as it has been shown explicitly in
Sec.~\ref{Phikstar} in the example of $(K^{\ast 0}+\bar{K}^{\ast
0})/2$ spectra (Fig.~\ref{Fig.11}). This is the case of the fits
to identified hadron spectra, since the statistical parameters
were fitted to the sample comprising 6 particle yields from PHENIX
($\pi^{\pm}$, $K^{\pm}$, $p$ and $\bar{p}$) and only 2 yield
ratios from STAR (see Ref.~\cite{Rafelski:2004dp}). And with these
parameters entering the expression for the invariant distribution,
Eq.~(\ref{Cooper2}), $p_{T}$ spectra of $\pi^{\pm}$, $K^{\pm}$,
$p$ and $\bar{p}$ measured by PHENIX have been fitted to determine
the geometric parameters of the model. Thus the main results
presented in Table~\ref{Table1} have been obtained within
practically one experiment, \textit{i.e.} the PHENIX
Collaboration.

The above remark should be kept in mind when the SHM predictions
for yields of other particles (other then used in the fitting
procedure) are compared with the data. For instance, in
Ref.~\cite{Letessier:2005kc} predictions for (anti)hyperons were
done on the basis of fits from Ref.~\cite{Rafelski:2004dp} (that
is fits done to the data set of which the main part is from PHENIX
at $\sqrt{s_{NN}}=200$ GeV). However the conclusion is drawn from
the comparison with the STAR data at $\sqrt{s_{NN}}=130$ GeV. The
conclusion is that the chemical full non-equilibrium case is in
favor. But in the main figure of Ref.~\cite{Letessier:2005kc}
(Fig.2 there), which has led to this conclusion, there are no
corresponding predictions in the strangeness chemical
non-equilibrium case. It is only stated there that these
predictions are in between the chemical full non-equilibrium and
the chemical equilibrium cases. All above arguments suggest that
the strangeness chemical non-equilibrium case as well as the
chemical equilibrium case have not been discredited entirely in
the context of the (anti)hyperon production.

Also particle yield fluctuations has been proposed as a definite
test of what scenario of the SHM is the most likely
\cite{Torrieri:2005va} but one has to await for the appropriate
data to make a conclusion. This test distinguishes between
(semi)equilibrium and non-equilibrium scenarios. But what is
interesting, values of the statistical parameters given there for
the chemical full non-equilibrium case are again at the condition
for the Bose-Einstein condensation of neutral pions, $\gamma_{q}
\approx \gamma_{q}^{cr}$, (these values are $T = 140.0$ MeV and
$\gamma_{q}=1.62$ \cite{Torrieri:2005va}, which gives $\mu_{\pi} =
135.079$ MeV, Eq.~(\ref{Pichempot}), so $\mu_{\pi} > m_{\pi^{0}}$
but this is the matter of rounding off \cite{Rafelski:2006pc}, if
one takes $\gamma_{q}=1.619$ then $\mu_{\pi} < m_{\pi^{0}}$). In
fact, as it is explained in
Refs.~\cite{Rafelski:2003ju,Letessier:2005qe}, $\gamma_{q}^{cr}$
is the upper limit of the allowed range of $\gamma_{q}$
superimposed before the fitting procedure has started. So by
definition $\gamma_{q} \leq \gamma_{q}^{cr}$ always (if the value
of $\gamma_{q}$ put in a table of
Refs.~\cite{Rafelski:2004dp,Rafelski:2003ju,Letessier:2005qe}
happens to exceed $\gamma_{q}^{cr}$ this is the result of rounding
up \cite{Rafelski:2006pc}, as in the above-mentioned example). But
from the technical point of view, when $\gamma_{q}$ slightly
exceeds this limit the fitting procedure will still proceed, since
$\pi^{0}$ yield is not included in the set of yields and/or ratios
to fit. So the true upper limit should be $e^{m_{\pi^{\pm}}/2T}$
rather, because exceeding this limit causes divergences in
primordial densities of $\pi^{+}$ and $\pi^{-}$, yields of which
are included usually in the set of data to fit. Thus the fitted
values of $\gamma_{q}$
\cite{Rafelski:2004dp,Rafelski:2003ju,Letessier:2005qe} seem to be
not trustworthy. All these facts put in question the idea of
introducing the parameter $\gamma_{q}$ into the model. But this
supports the conclusion that the chemical full non-equilibrium
freeze-out is the least likely. Anyway, if values of $\gamma_{q}$
were at the critical point for the Bose-Einstein condensation of
$\pi^{0}$ (as it is claimed in
Refs.~\cite{Rafelski:2004dp,Rafelski:2003ju,Letessier:2005qe}),
then the significant $\pi^{0}$-overproduction at low-$p_{T}$ could
happen with respect to one half of $(\pi^{+}+\pi^{-})$ (see
Sec.~\ref{Pizero}), which seems to be checkable at least in
principle.

And the last remark is that the present analysis has been done
within a particular hypersurface, as given by
Eqs.~(\ref{Hypsur})-(\ref{Transsiz}). Of course, the natural
question is to what extend the results depend on the choice of a
hypersurface. One of the indirect arguments pro this hypersurface
are the results of fits to the PHENIX spectra of $\pi^{\pm}$,
$K^{\pm}$, $p$ and $\bar{p}$ done in Ref.~\cite{Adler:2004hv}
within the very popular blast-wave model
\cite{Schnedermann:1993ws}. For those fits $\chi^{2}$/NDF$\approx
3-4$, so from the statistical point of view such a hypothesis
should be rejected. On the opposite, in the present work
$\chi^{2}$/NDF$<1$ has been obtained for all central and
mid-central bins (see Table~\ref{Table1}). Thus at least for these
bins the hypothesis that the hypersurface has the form as given
here can not be rejected. Also the discussed possible
overproduction of low-$p_{T}$ $\pi^{0}$'s seems not to depend very
much on a form of the hypersurface chosen since this effect is the
result of approaching the condition for the Bose-Einstein
condensation of neutral pions. This causes the abrupt increase of
the distribution function of $\pi^{0}$,
$f_{\pi^{0}}^{primordial}(p \cdot u)$, when $p_{T} \rightarrow 0$
for all hypersurfaces which have a region with the negligible
flow.

To summarize, in the view of this analysis the chemical full
non-equilibrium  freeze-out seems to happen least likely during
Au-Au collisions at $\sqrt{s_{NN}}=200$ GeV and both other cases
are of the similar likelihood. To help to verify this conclusion,
the low-$p_{T}$ $\pi^{0}$ measurement is proposed since at low
$p_{T}$ the ratio of $\pi^{0}$ over one half of
$(\pi^{+}+\pi^{-})$ distinguishes very clearly between $\gamma_{q}
\approx \gamma_{q}^{cr}$ and $\gamma_{q}=1$.

\begin{acknowledgments}
The author would like to thank Jan Rafelski and Wojciech
Florkowski for very helpful discussions. This work was supported
in part by the Polish Committee for Scientific Research under
Contract No. KBN 2 P03B 069 25.
\end{acknowledgments}


\begin{thebibliography}{99}
\bibitem{Rafelski:2004dp}
J.~Rafelski, J.~Letessier and G.~Torrieri,
Phys.\ Rev.\ C {\bf 72}, 024905 (2005).

\bibitem{Adler:2003cb}
S.~S.~Adler {\it et al.}  [PHENIX Collaboration],
Phys.\ Rev.\ C {\bf 69}, 034909 (2004).

\bibitem{Letessier:2002gp}
J.~Letessier and J.~Rafelski,
Cambridge Monogr. Part. Phys. Nucl. Phys. Cosmol. {\bf 18} 1
(2002).

\bibitem{Zhang:2004rj}
H.~B.~Zhang  [STAR Collaboration],
arXiv:nucl-ex/0403010.

\bibitem{Adams:2004ep}
J.~Adams {\it et al.}  [STAR Collaboration],
Phys.\ Rev.\ C {\bf 71}, 064902 (2005).

\bibitem{Adams:2004ux}
J.~Adams {\it et al.}  [STAR Collaboration],
Phys.\ Lett.\ B {\bf 612}, 181 (2005).

\bibitem{Prorok:2005uv}
D.~Prorok,
Phys.\ Rev.\ C {\bf 73}, 064901 (2006).

\bibitem{Broniowski:2001we}
W.~Broniowski and W.~Florkowski,
Phys.\ Rev.\ Lett.\  {\bf 87}, 272302 (2001).

\bibitem{Broniowski:2001uk}
W.~Broniowski and W.~Florkowski,
Phys.\ Rev.\ C {\bf 65}, 064905 (2002).

\bibitem{Broniowski:2002nf}
W.~Broniowski, A.~Baran and W.~Florkowski,
Acta Phys.\ Polon.\ B {\bf 33}, 4235 (2002).

\bibitem{Adler:2004zn}
S.~S.~Adler {\it et al.}  [PHENIX Collaboration],
Phys.\ Rev.\ C {\bf 71}, 034908 (2005) [Erratum-ibid.\ C {\bf 71},
049901 (2005)].

\bibitem{Back:2006yw}
B.~B.~Back {\it et al.}  [PHOBOS Collaboration],
Phys.\ Rev.\ C {\bf 74}, 021902(R) (2006).

\bibitem{Hagiwara:fs}
K.~Hagiwara {\it et al.}  [Particle Data Group Collaboration],
Phys.\ Rev.\ D {\bf 66}, 010001 (2002).

\bibitem{Prorok:2004af}
D.~Prorok,
Eur.\ Phys.\ J.\ A {\bf 24}, 93 (2005).

\bibitem{Adcox:2001ry}
K.~Adcox {\it et al.}  [PHENIX Collaboration],
Phys.\ Rev.\ Lett.\  {\bf 87}, 052301 (2001).

\bibitem{Bearden:2003hx}
I.~G.~Bearden {\it et al.}  [BRAHMS Collaboration],
Phys.\ Rev.\ Lett.\  {\bf 93}, 102301 (2004).

\bibitem{Adler:2004hv}
S.~S.~Adler {\it et al.}  [PHENIX Collaboration],
Phys.\ Rev.\ C {\bf 72}, 014903 (2005).

\bibitem{Back:2004zx}
B.~B.~Back  {\it et al.}  [PHOBOS Collaboration],
Phys.\ Rev.\ C {\bf 70}, 051901(R) (2004).

\bibitem{Barannikova:2005rw}
O.~Barannikova  [STAR Collaboration],
J.\ Phys.\ G {\bf 31}, S93 (2005).

\bibitem{Adams:2003xp}
J.~Adams {\it et al.}  [STAR Collaboration],
Phys.\ Rev.\ Lett.\  {\bf 92}, 112301 (2004).

\bibitem{Adler:2003qi}
S.~S.~Adler {\it et al.}  [PHENIX Collaboration],
Phys.\ Rev.\ Lett.\  {\bf 91}, 072301 (2003).

\bibitem{Rafelski:2003ju}
J.~Rafelski and J.~Letessier,
Acta Phys.\ Polon.\ B {\bf 34}, 5791 (2003).

\bibitem{Letessier:2005qe}
J.~Letessier and J.~Rafelski,
arXiv:nucl-th/0504028.

\bibitem{Chujo:2002bi}
T.~Chujo  [PHENIX Collaboration],
Nucl.\ Phys.\ A {\bf 715}, 151 (2003) and \newline
\texttt{http://alice-france.in2p3.fr/qm2002/Transparencies/20Plenary/Chujo.ppt}.

\bibitem{Adcox:2001mf}
K.~Adcox {\it et al.}  [PHENIX Collaboration],
Phys.\ Rev.\ Lett.\  {\bf 88}, 242301 (2002).

\bibitem{Letessier:2005kc}
J.~Letessier and J.~Rafelski,
Phys.\ Rev.\ C {\bf 73}, 014902 (2006).

\bibitem{Torrieri:2005va}
G.~Torrieri, S.~Jeon and J.~Rafelski,
Phys.\ Rev.\ C {\bf 74}, 024901 (2006).

\bibitem{Rafelski:2006pc}
J.~Rafelski (private communication).

\bibitem{Schnedermann:1993ws}
E.~Schnedermann, J.~Sollfrank and U.~Heinz,
Phys.\ Rev.\ C {\bf 48}, 2462 (1993).
\end{thebibliography}
\end{document}